\documentclass[]{pasj02}

\usepackage[switch,mathlines]{lineno} 

\jyear{2025}

\Received{2025/01/16}
\Accepted{2025/05/05}

\newcommand{\Ni}{\ensuremath{^{56}\mathrm{Ni}}\xspace}

\newcommand{\Msun}{\,\ensuremath{\mathrm{M}_\odot}\xspace}
\newcommand{\Rsun}{\,\ensuremath{\mathrm{R}_\odot}\xspace}
\newcommand{\Zsun}{\,\ensuremath{\mathrm{Z}_\odot}\xspace}
\newcommand{\Mzams}{\,\ensuremath{M_\mathrm{ZAMS}}\xspace}
\newcommand{\Msunpyr}{\,\ensuremath{\Msun~\mathrm{yr^{-1}}}\xspace}
\newcommand{\kmps}{\,\ensuremath{\mathrm{km~s^{-1}}}\xspace}

\graphicspath{{./}{figures/}}

\usepackage{xspace}
\usepackage{natbib}
\usepackage{url}
 
\begin{document} 

\title{ 
Properties of high-redshift Type~II supernovae discovered by the JADES transient survey
}

\author{
 Takashi J. \textsc{Moriya},\altaffilmark{1,2,3}\altemailmark\orcid{0000-0003-1169-1954} \email{takashi.moriya@nao.ac.jp} 
 David A. \textsc{Coulter},\altaffilmark{4}\orcid{0000-0003-4263-2228}
 Christa \textsc{DeCoursey},\altaffilmark{5}\orcid{0000-0002-4781-9078}
 Justin D. R. \textsc{Pierel},\altaffilmark{4}\orcid{0000-0002-2361-7201}
 Kevin \textsc{Hainline},\altaffilmark{5}\orcid{0000-0003-4565-8239}
 Matthew R. \textsc{Siebert},\altaffilmark{4}\orcid{0000-0003-2445-3891}
 Armin \textsc{Rest},\altaffilmark{4,6}\orcid{0000-0002-4410-5387} 
 Eiichi \textsc{Egami},\altaffilmark{5}\orcid{0000-0003-1344-9475} 
 Sebastian \textsc{Gomez},\altaffilmark{7}\orcid{0000-0001-6395-6702}
 Robert M. \textsc{Quimby},\altaffilmark{8,9}\orcid{0000-0001-9171-5236}
 Ori D. \textsc{Fox},\altaffilmark{4}\orcid{0000-0003-2238-1572}
 Michael \textsc{Engesser},\altaffilmark{4}\orcid{0000-0003-0209-674X}
 Fengwu \textsc{Sun},\altaffilmark{7}\orcid{0000-0002-4622-6617}
 Wenlei \textsc{Chen},\altaffilmark{10}\orcid{0000-0003-1060-0723}
 Yossef \textsc{Zenati},\altaffilmark{6,4}\orcid{0000-0002-0632-8897}
 Suvi \textsc{Gezari},\altaffilmark{4,6}\orcid{0000-0003-3703-5154}
 Bhavin A. \textsc{Joshi},\altaffilmark{6}\orcid{0000-0002-7593-8584}
 Melissa \textsc{Shahbandeh},\altaffilmark{4}\orcid{0000-0002-9301-5302}
 Louis-Gregory \textsc{Strolger},\altaffilmark{4}\orcid{0000-0003-2238-1572}
 Qinan \textsc{Wang},\altaffilmark{11}\orcid{0000-0001-5233-6989} 
 Stacey \textsc{Alberts},\altaffilmark{5}\orcid{0000-0002-8909-8782}
 Rachana  \textsc{Bhatawdekar},\altaffilmark{12}\orcid{0000-0003-0883-2226}
 Andrew J. \textsc{Bunker},\altaffilmark{13}\orcid{0000-0002-8651-9879}
 Pierluigi \textsc{Rinaldi},\altaffilmark{5}\orcid{0000-0002-5104-8245}
 Brant E. \textsc{Robertson},\altaffilmark{14}\orcid{0000-0002-4271-0364}
 and 
 Sandro \textsc{Tacchella}\altaffilmark{15,16}\orcid{0000-0002-8224-4505}
}
\altaffiltext{1}{National Astronomical Observatory of Japan, National Institutes of Natural Sciences, 2-21-1 Osawa, Mitaka, Tokyo 181-8588, Japan}
\altaffiltext{2}{Graduate Institute for Advanced Studies, SOKENDAI, 2-21-1 Osawa, Mitaka, Tokyo 181-8588, Japan}
\altaffiltext{3}{School of Physics and Astronomy, Monash University, Clayton, VIC 3800, Australia}
\altaffiltext{4}{Space Telescope Science Institute, Baltimore, MD 21218, USA}
\altaffiltext{5}{Steward Observatory, University of Arizona, 933 N. Cherry Avenue, Tucson, AZ 85721, USA}
\altaffiltext{6}{Physics and Astronomy Department, Johns Hopkins University, Baltimore, MD 21218, USA}
\altaffiltext{7}{Center for Astrophysics $|$ Harvard \& Smithsonian, 60 Garden Street, Cambridge, MA 02138-1516, USA}
\altaffiltext{8}{Department of Astronomy/Mount Laguna Observatory, San Diego State University, 5500 Campanile Drive, San Diego, CA 92812-1221, USA}
\altaffiltext{9}{Kavli Institute for the Physics and Mathematics of the Universe (WPI), The University of Tokyo Institutes for Advanced Study, The University of Tokyo, Kashiwa, Chiba 277-8583, Japan}
\altaffiltext{10}{Department of Physics, Oklahoma State University, 145 Physical Sciences Bldg, Stillwater, OK 74078, USA}
\altaffiltext{11}{Department of Physics and Kavli Institute for Astrophysics and Space Research, Massachusetts Institute of Technology, 77 Massachusetts Avenue, Cambridge, MA 02139, USA}
\altaffiltext{12}{European Space Agency (ESA), European Space Astronomy Centre (ESAC), Camino Bajo del Castillo s/n, 28692 Villanueva de la Cañada, Madrid, Spain}
\altaffiltext{13}{Department of Physics, University of Oxford, Denys Wilkinson Building, Keble Road, Oxford OX1 3RH, UK}
\altaffiltext{14}{Department of Astronomy and Astrophysics, University of California, Santa Cruz, 1156 High Street, Santa Cruz CA 96054, USA}
\altaffiltext{15}{Kavli Institute for Cosmology, University of Cambridge, Madingley Road, Cambridge, CB3 0HA, UK}
\altaffiltext{16}{Cavendish Laboratory, University of Cambridge, 19 JJ Thomson Avenue, Cambridge, CB3 0HE, UK}

\KeyWords{supernovae: general --- supernovae: individual (AT~2023adsv, AT~2023adte, AT~2023adtf, SN~2023adto, SN~2023adtu, AT~2023adtw) --- stars: massive}

\maketitle

\begin{abstract}
In this work we estimate the explosion and progenitor properties of six Type~II supernovae (SNe) at $0.675\leq z\leq 3.61$ discovered by the James Webb Space Telescope (JWST) Advanced Deep Extragalactic Survey (JADES) transient survey by modeling their light curves.
Two Type~II SNe are found to have high explosion energies of $3\times 10^{51}~\mathrm{erg}$, while the other four Type~II SNe are estimated to have typical explosion energies found in the local Universe [$(0.5-2)\times 10^{51}~\mathrm{erg}$]. The fraction of Type~II SNe with high explosion energies might be higher at high redshifts because of, e.g., lower metallicity, but it is still difficult to draw a firm conclusion because of the small sample size and potential observational biases. We found it difficult to constrain the progenitor masses for Type~II SNe in our sample because of the sparse light-curve data. We found two Type~II SN light curves can be better reproduced by introducing confined, dense circumstellar matter. Thus, the confined, dense circumstellar matter frequently observed in nearby Type~II SNe is likely to exist in Type~II SNe at high redshifts as well. Two Type~II SNe are estimated to have high host galaxy extinctions, showing the ability of JWST to discover dust-obscured SNe at high redshifts. More high-redshift Type~II SNe are required to investigate the differences in the properties of Type~II SNe near and far, but here we show the first glimpse into the high-redshift population of Type~II SNe.
\end{abstract}


\begin{longtable}{lccccc}
  \caption{Type~II SN properties analyzed in this work.}\label{tab:sample}
  \hline              
  Name & Redshift & $E(B-V)$\footnotemark[$*$] & ZAMS mass\footnotemark[$\dag$] & Explosion energy & Mass-loss rate\footnotemark[$\ddag$]  \\ 
   &  & ($\mathrm{mag}$) & (\Msun) & ($\mathrm{B}$) & (\Msunpyr)  \\  
\endfirsthead
  \hline
  Name & Redshift & Extinction & ZAMS mass & Explosion energy & Mass-loss rate  \\
\endhead
  \hline
\endfoot
  \hline
  \multicolumn{6}{l}{\footnotemark[$*$] Extinction estimated from light-curve modeling. }\\
  \multicolumn{6}{l}{\footnotemark[$\dag$] The ZAMS masses adopted in this paper is $12, 16, 20,$ and 24~\Msun.}\\
  \multicolumn{6}{l}{\footnotemark[$\ddag$] The mass-loss rate forming the confined dense CSM assuming $v_\infty=10~\kmps$. The CSM radius is $10^{15}~\mathrm{cm}$.}\\
  \multicolumn{6}{l}{\footnotemark[$\S$] Photometrically classified as a Type~II SN by \citet{decoursey2024}.}\\
  \multicolumn{6}{l}{\footnotemark[$\|$] Spectroscopically classified as a Type~II SN by Egami et al. (in preparation).}\\
  \multicolumn{6}{l}{\footnotemark[$\sharp$] Host galaxy spectroscopic redshift from \citet{coulter2024}.}\\
  \multicolumn{6}{l}{\footnotemark[$**$] Host galaxy spectroscopic redshift from \citet{deugenio2024}.}\\
  \multicolumn{6}{l}{\footnotemark[$\dag\dag$] Host galaxy spectroscopic redshift from \citet{bunker2023}.}\\
  \multicolumn{6}{l}{\footnotemark[$\ddag\ddag$] SN spectroscopic redshift from Egami et al. (in preparation).}\\
  \multicolumn{6}{l}{\footnotemark[$\S\S$] Host galaxy spectroscopic redshift from \citet{momcheva2016}.}\\
\endlastfoot
  \hline
  AT~2023adsv\footnotemark[$\S$] & $3.61 $\footnotemark[$\sharp$] & $0$ & $20-24$ & $3.0$ & $10^{-3}$ \\
  AT~2023adte\footnotemark[$\S$] & $2.623$\footnotemark[$**$] & $0$ & $16-24$ & $1.5$ & -  \\
  AT~2023adtf\footnotemark[$\S$] & $2.344$\footnotemark[$\dag\dag$] & $0$ & $12-24$ & $0.5-0.6$ & $1.3\times 10^{-3}$ \\
  SN~2023adto\footnotemark[$\|$] & $1.62 $\footnotemark[$\ddag\ddag$] & $0$ & $12-24$ & $1.2-1.7$ & -  \\ 
  SN~2023adtu\footnotemark[$\|$] & $1.01 $\footnotemark[$\ddag\ddag$] & $0.25$ & $12-16$ & $1.5-2.3$ & -  \\
  AT~2023adtw\footnotemark[$\S$] & $0.657$\footnotemark[$\S\S$] & $1.3-1.5$ & $12-24$ & $3.0$ & -  \\
\end{longtable}

\begin{table}
  \tbl{SN progenitor properties.}{%
  \begin{tabular}{cccc}
      \hline
      $M_\mathrm{ZAMS}$ & $M_\mathrm{total}$\footnotemark[$*$] & $M_\mathrm{H-rich}$\footnotemark[$\dag$] & $R$\footnotemark[$\ddag$] \\
      \hline
  $12~\Msun$ & $11.9~\Msun$ &  $8.7~\Msun$ & $394~\Rsun$  \\
  $16~\Msun$ & $15.7~\Msun$ & $10.7~\Msun$ & $593~\Rsun$  \\
  $20~\Msun$ & $18.4~\Msun$ & $11.6~\Msun$ & $762~\Rsun$  \\
  $24~\Msun$ & $18.9~\Msun$ &  $8.9~\Msun$ & $996~\Rsun$  \\ 
      \hline
    \end{tabular}}\label{tab:progenitors}
\begin{tabnote}
\footnotemark[$*$] Total mass at explosion  \\ 
\footnotemark[$\dag$] Hydrogen-rich envelope mass at explosion \\
\footnotemark[$\ddag$] Radius at explosion \\
\end{tabnote}
\end{table}

\section{Introduction}
Massive stars play fundamental roles in the cosmic history (\citealt{eldridge2022} for a recent review). For example, strong radiation is released from massive stars throughout their evolution and they affect the ionization states of the surrounding environments \citep[e.g.,][]{rix2004}. Massive stars die as supernovae (SNe) at the end of their evolution. SNe provide metals and energies to the surroundings, driving the chemical and dynamical evolution of the Universe \citep[e.g.,][]{nomoto2013}. Understanding massive star evolution and resulting SNe across cosmic time is essential to reveal the history of the Universe.

Type~II SNe consist of around 70\% of core-collapse SNe \citep[e.g.,][]{li2011,shivvers2017}. Thus, understanding Type~II SNe is important to reveal the general properties of core-collapse SNe. Revealing the general properties of Type~II SNe is also essential to uncovering the standard explosion mechanisms of core-collapse SNe \citep[e.g.,][]{muller2019,burrows2021,nakamura2024}. Hundreds of Type~II SNe have been discovered so far and their general properties have been investigated by many studies \citep[e.g.,][]{martinez2022,subrayan2023,silva-farfan2024}. However, Type~II SN discoveries are so far mostly limited to the local Universe at $z<0.4$ \citep{dejaeger2017,dejaeger2020,gall2018} except for a couple of photometrically identified Type~II SNe at $z\sim 2$ from the Cosmic Assembly Near-infrared Deep Extragalactic Legacy Survey (CANDELS, \citealt{rodney2014,strolger2015}). Type~II SN properties at high redshifts can be different from those in the local Universe because of, e.g., differences in metallicity \citep{dessart2014,anderson2018}. In addition to being tracers of stellar explosion properties, high-redshift Type~II SNe can be useful as a distance indicator \citep{dejaeger2020}.

Discovering high-redshift Type~II SNe to constrain their properties is challenging. Optical transient surveys have been discovering SNe at high redshifts, but they have been mostly luminous SNe like superluminous SNe \citep[e.g.,][]{cooke2012,pan2017,smith2018,moriya2019,curtin2019}. Recently, however, \citet{decoursey2024} reported the discovery of 79 SNe at $0.21\leq z\leq 4.82$ from the JWST Advanced Deep Extragalactic Survey (JADES, \citealt{eisenstein2023}) transient survey. The JADES transient survey is conducted in one of the JADES Deep Fields located in the Great Observatories Origins Deep Survey (GOODS) South Field \citep[e.g.,][]{giavalisco2004}. Deep JWST imaging observations reaching $\sim 30~\mathrm{mag}$ ($5\sigma$) were conducted in the $\sim 25~\mathrm{arcmin^2}$ survey field for five epochs spanning over a year and many transient objects have been identified \citep{decoursey2024}. Follow-up JWST spectroscopic observations were also triggered to confirm the nature of the transients (Egami et al. in preparation). Thanks to the deep JWST data, they discovered SNe with typical brightness such as a Type~Ia SN at $z=2.9$ \citep{pierel2024} and a broad-lined Type~Ic SN at $z=2.83$ \citep{siebert2024}. 

Among 79 SNe, 46 SNe are classified as Type~II either photometrically or spectroscopically (\citealt{decoursey2024}; Egami et al. in preparation). Among them, we take six Type~II SNe that are spectroscopically confirmed as Type~II SNe or that are photometrically confirmed as Type~II SNe with decent light-curve information and the confirmed redshift from their host galaxy spectra. The highest redshift Type~II SN with the spectroscopic redshift is AT~2023adsv at $z=3.61$, which is discussed in more detail in a separate paper \citep{coulter2024}. We estimate the high-redshift Type~II SN properties by modeling their light curves and discuss if there are any possible differences in the properties of Type~II SNe near and far. This high-redshift Type~II SN sample allows us to investigate the properties of typical Type~II SNe at high redshifts for the first time. 

The rest of this paper is organized as follows. We first introduce our six high-redshift Type~II SNe used to estimate their properties in Section~\ref{sec:sample}. We introduce the low-metallicity red supergiant (RSG) SN progenitor models adopted in this work in Section~\ref{sec:progenitor}. The method of the light-curve modeling is introduced in Section~\ref{sec:lightcurve}. Our method of host galaxy spectral energy distribution (SED) fitting from which we estimate the host galaxy attenuation is described in Section~\ref{sec:hostfitting}. We show the results of our estimations of Type~II SN properties in Section~\ref{sec:properties}. We discuss the possible differences in the Type~II SN properties at high and low redshifts and conclude this paper in Section~\ref{sec:discussion}. A standard $\Lambda$CDM cosmology with $H_0=70~\mathrm{km~s^{-1}~Mpc^{-1}}$, $\Omega_M=0.3$, and $\Omega_\Lambda = 0.7$ is adopted. The AB magnitude system is used in this paper.

\section{Methods}\label{sec:methods}
\subsection{Type~II SN selection}\label{sec:sample}
Among the 46 Type~II SNe identified by \citet{decoursey2024}, we chose 6 Type~II SNe to perform the numerical modeling to estimate their properties with the following criteria. First, we take two spectroscopically confirmed Type~II SNe (SN~2023adto and SN~2023adtu). 
The spectra of SN~2023adto and SN~2023adtu show clear P-Cygni profiles of H$\alpha$ and Ca~\textsc{ii} 8600~\AA\ triplet and match well to a Type~II SN spectral template from SN~1999em \citep{elmhamdi2003}. The redshifts to these SNe were determined by the SN template matching as $z=1.62$ (SN~2023adto) and $z=1.01$ (SN~2023adtu). Although the redshift determination by the SN template matching is not as accurate as the redshift determination by host galaxy spectra because of, e.g., intrinsic diversities in line shifts observed in SNe \citep{anderson2014spec}, our results of light-curve modeling is not strongly affected by the redshift uncertainties as long as they are determined within a few percent. These spectra will be reported in Egami et al. (in preparation).

\begin{figure*}
 \begin{center}
  \includegraphics[width=5cm]{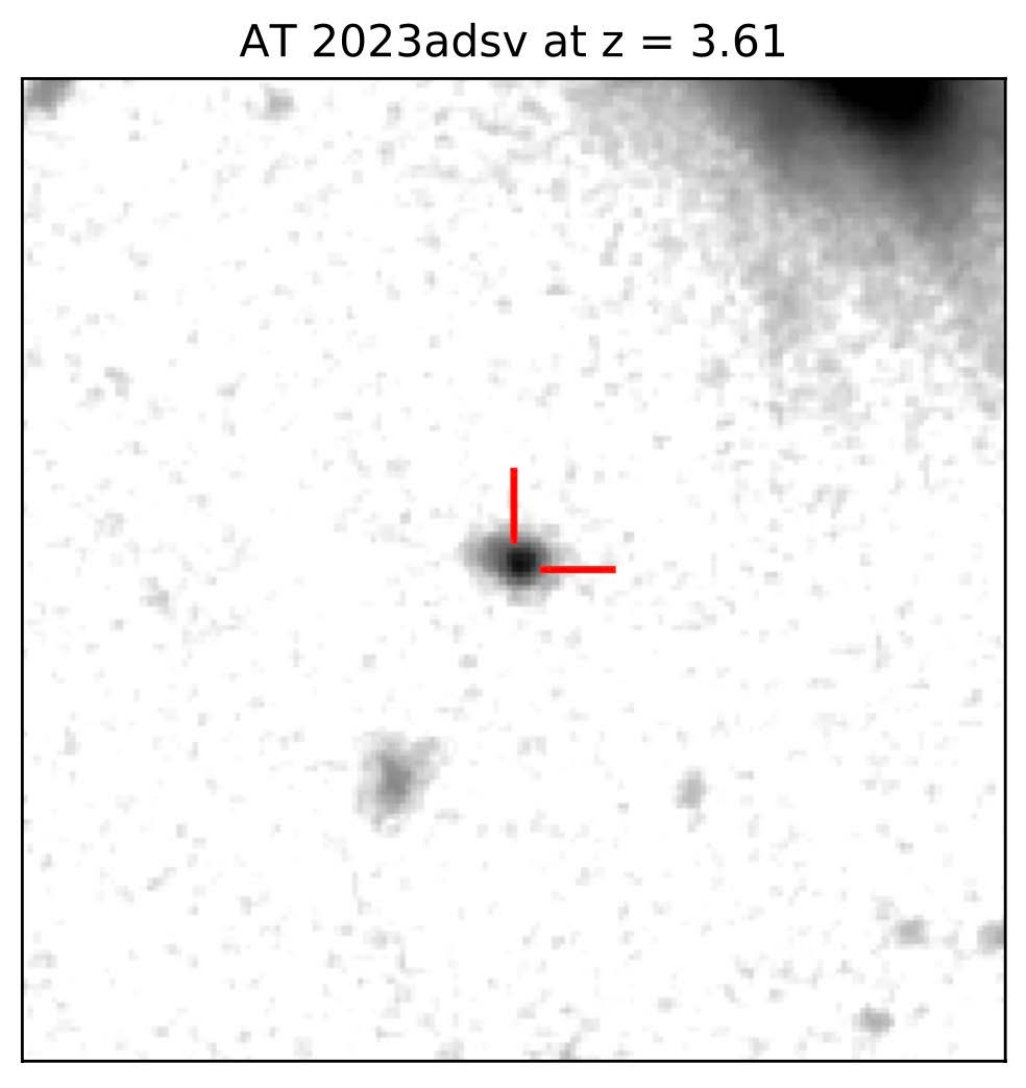} 
  \includegraphics[width=5cm]{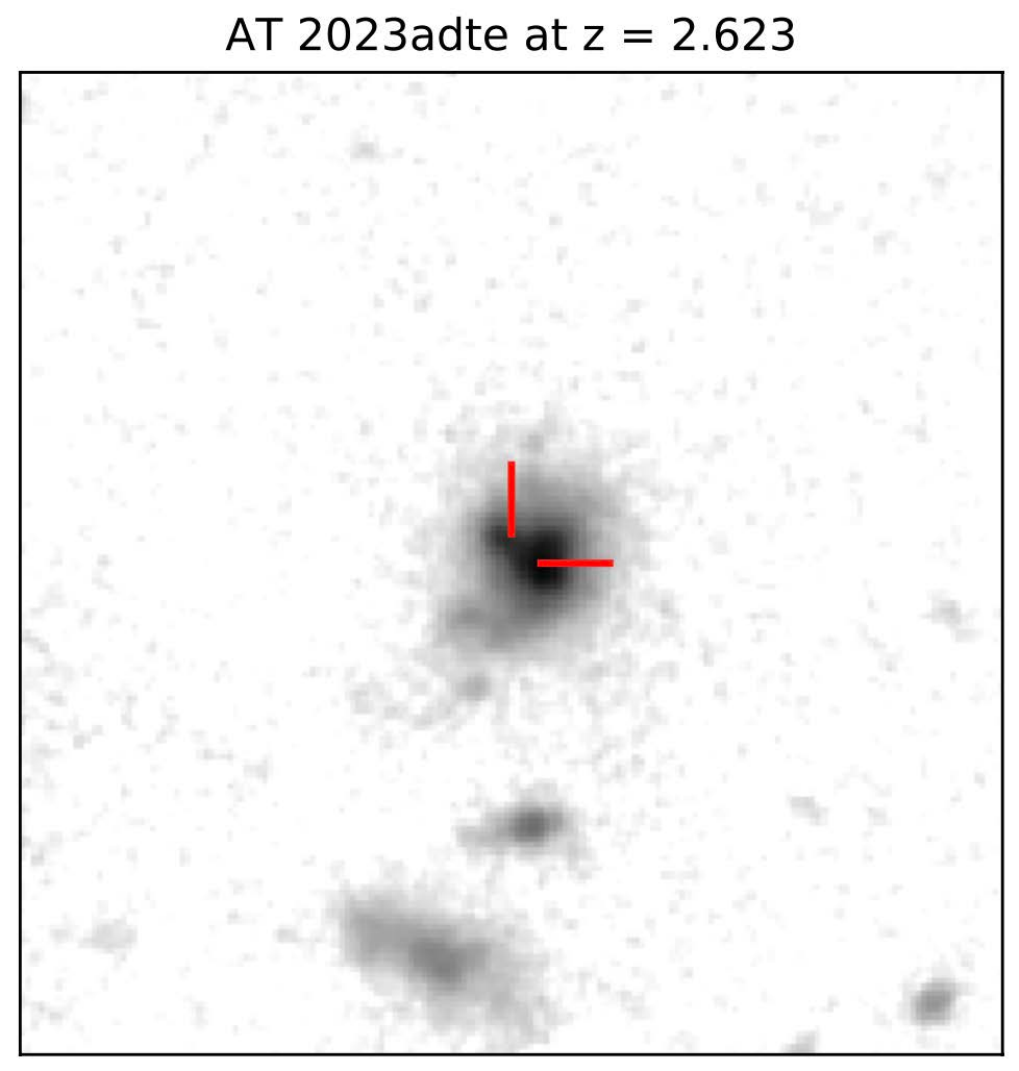} 
  \includegraphics[width=5cm]{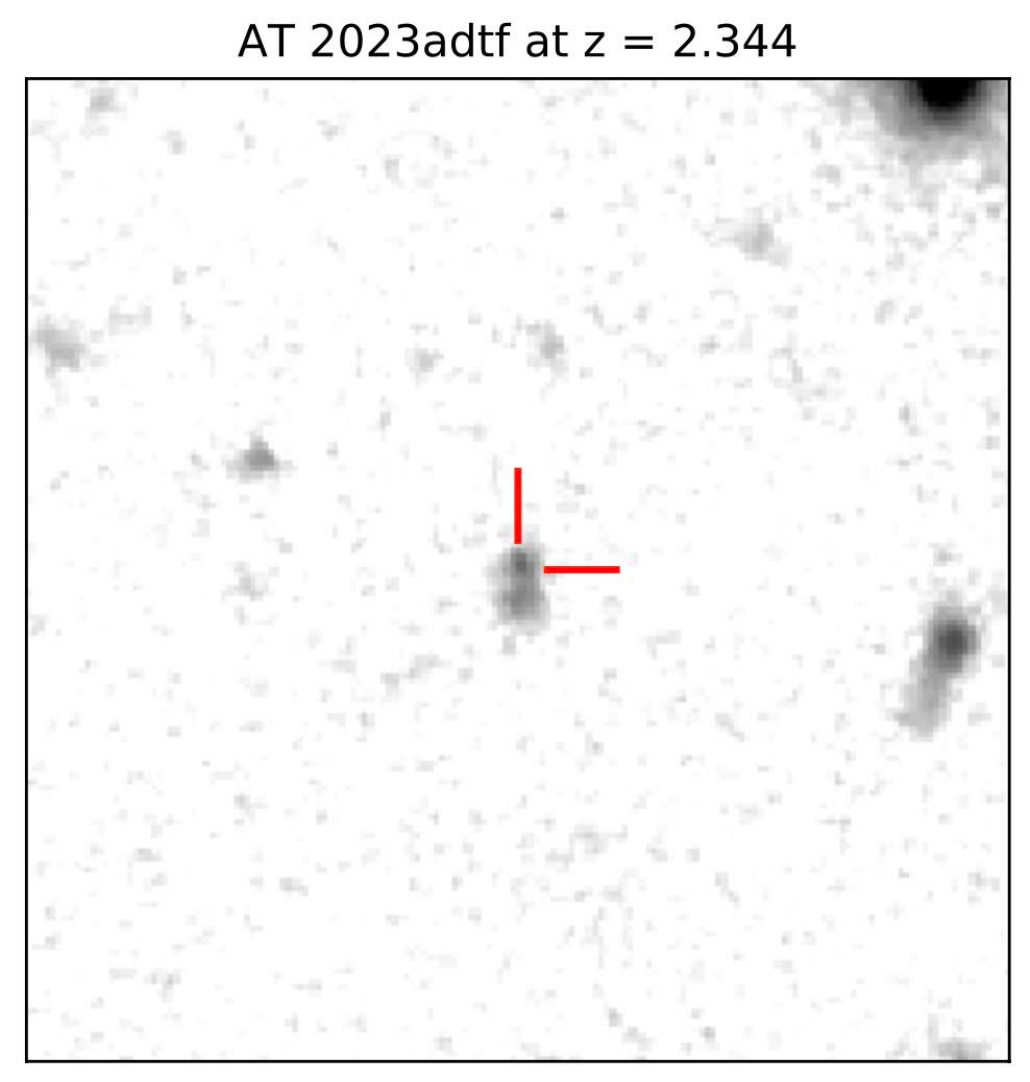} \\
  \includegraphics[width=5cm]{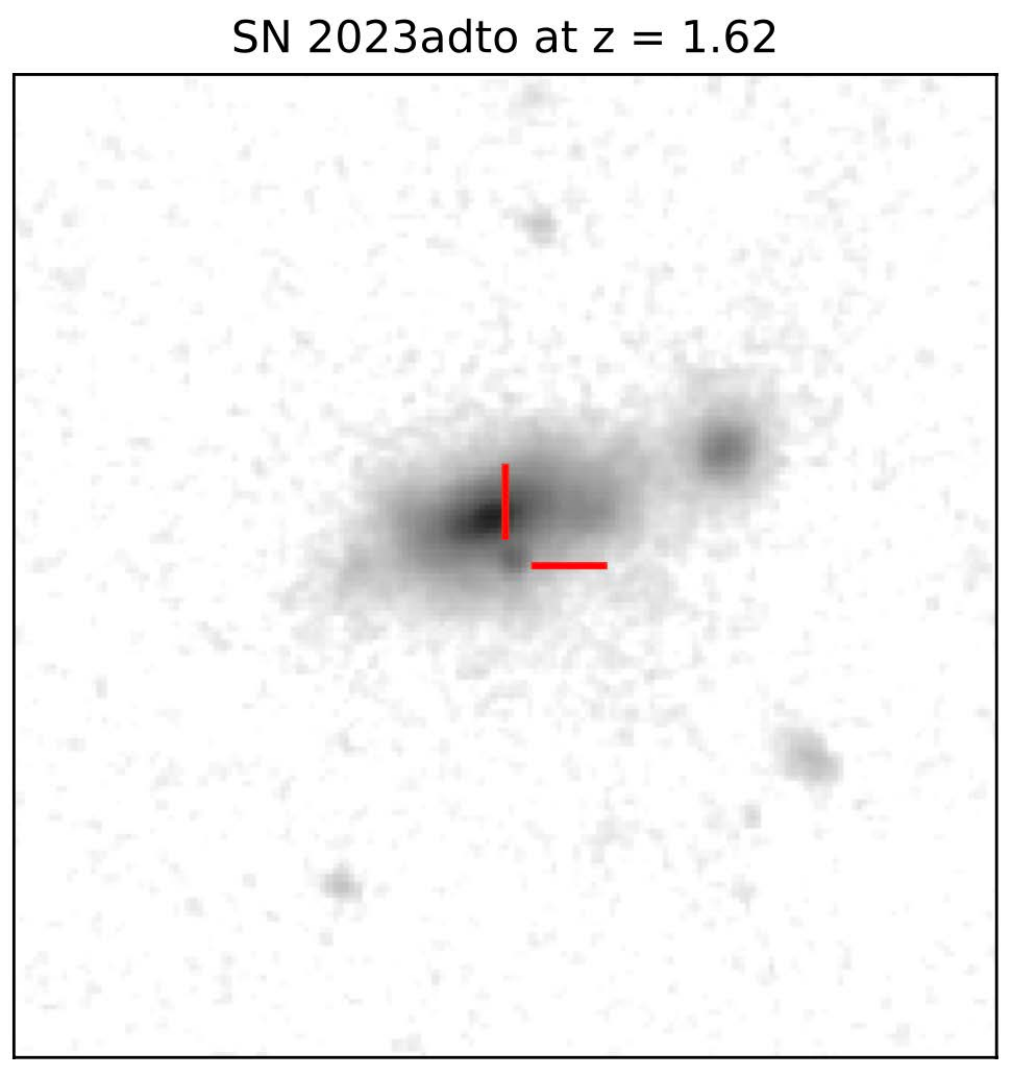}
  \includegraphics[width=5cm]{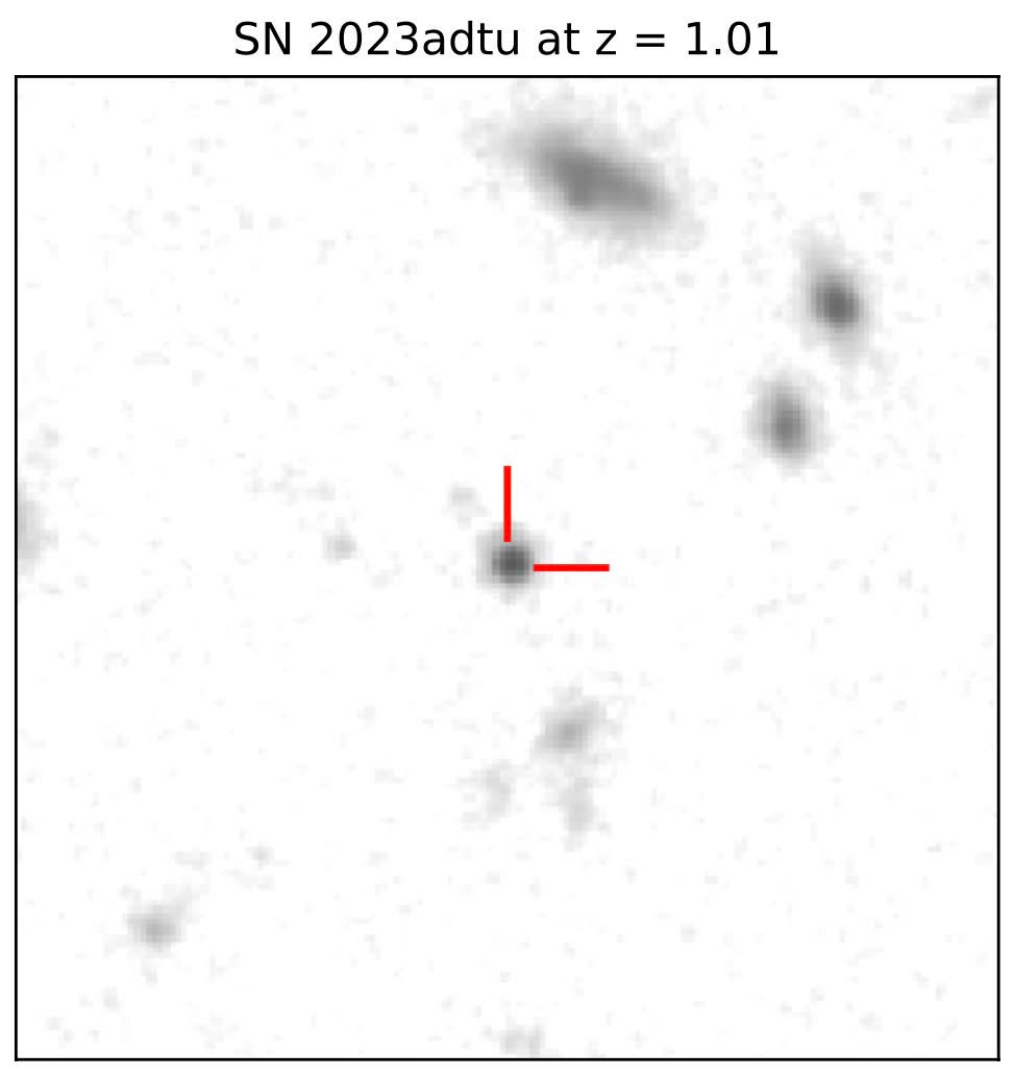}
  \includegraphics[width=5cm]{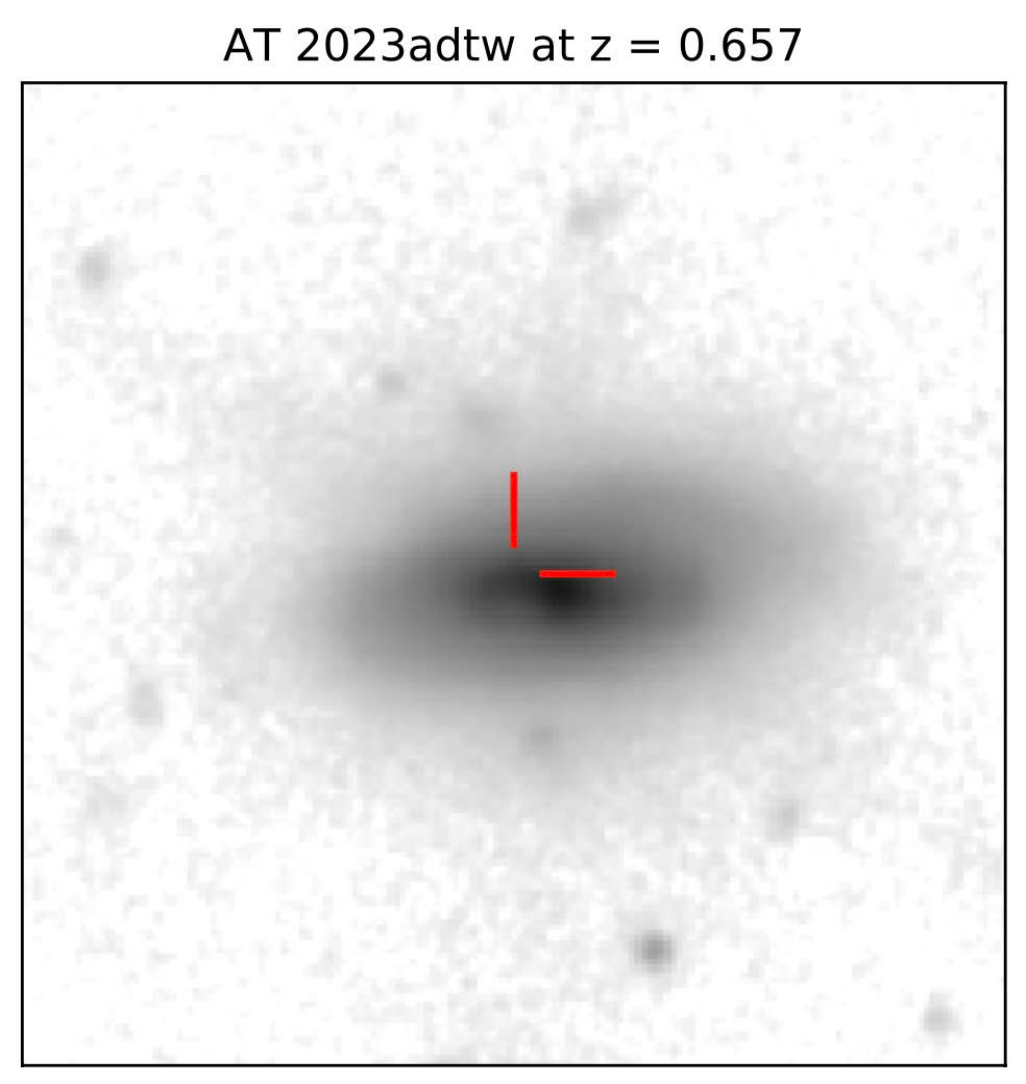}
 \end{center}
\caption{Images of the Type~II SNe discussed in this paper. They are 5"x5" images in the F277W filter. SNe are located at the center.
North is up and east is left. Reference and difference images, as well as images in the other filters, are available in \citet{decoursey2024}. {Alt text: Images showing flux strength. Darker regions have more flux.} 
} 
\label{fig:face}
\end{figure*}

For SNe without spectroscopic classification, we adopt the photometric classification by \citet{decoursey2024} which is based on the \texttt{STARDUST2} Bayesian light curve classification tool \citep{rodney2014}. We take SNe (1) showing 100\% probability to be Type~II SNe, (2) having spectroscopically confirmed redshifts from their host galaxies, and (3) observed at least for three epochs. Four SN candidates (AT~2023adsv, AT~2023adte, AT~2023adtf, and AT~2023adtw) satisfy the three criteria and we add them to our sample of Type~II SNe to analyze. Table~\ref{tab:sample} summarizes our sample of Type~II SNe. Their images are shown in Figure~\ref{fig:face}.

\subsection{Progenitor calculations}\label{sec:progenitor}
The redshift range of Type~II SNe we analyze in this paper is from $z=0.657$ to $z=3.61$. Because the average metallicities in this redshift range are of the order of $0.1~\Zsun$ \citep[e.g.,][]{curti2024}, we develop SN progenitor models with $Z = 0.1~\Zsun$ ($\Zsun = 0.014$, \citealt{asplund2009}) by using \texttt{Modules for Experiments in Stellar Astrophysics} (\texttt{MESA}, \citealt{paxton2011,paxton2013,paxton2015,paxton2018,paxton2019,jermyn2023}) version r23.05.1. We calculate the SN progenitors with the zero-age main-sequence (ZAMS) masses (\Mzams) of $12, 16, 20,$ and $24~\Msun$. We adopt the Schwarzschild criterion for convection with a mixing-length parameter of $2$. The stellar evolution calculation is performed beyond the core carbon depletion. The structure of the hydrogen-rich envelope, which determines the light-curve properties at the luminous phases in Type~II SNe, does not change much after the core carbon depletion. 

Mass-loss can be strongly affected by metallicity. We adopt the mass-loss prescription from \citet{vink2001} when the effective temperature ($T_\mathrm{eff}$) is above $14000~\mathrm{K}$. The mass-loss rate ($\dot{M}$) scales as $\dot{M}\propto Z^{0.69-0.64}$ depending on the effective temperature \citep{vink2001}. For the cool phase with $T_\mathrm{eff}<8000~\mathrm{K}$, we adopt the RSG mass-loss prescription of \citet{dejager1988}. Because the metallicity dependence of the RSG mass-loss is less significant (\citealt{yang2023,kee2021,goldman2017}, but see also \citealt{mauron2011}), we do not take the metallicity dependence of the RSG mass-loss into account. In the effective temperature range of $8000~\mathrm{K}<T_\mathrm{eff}<14000~\mathrm{K}$, the hot-star mass-loss rate at $T_\mathrm{eff}=14000~\mathrm{K}$ and the cool-star mass-loss rate at $T_\mathrm{eff}=8000~\mathrm{K}$ are interpolated.

All the progenitors ($M_\mathrm{ZAMS}=12, 16, 20,$ and $24~\Msun$) are RSGs at the core carbon depletion stage and they are Type~II SN progenitors. The progenitor properties are summarized in Table~\ref{tab:progenitors}.

\begin{figure}
 \begin{center}
  \includegraphics[width=7.5cm]{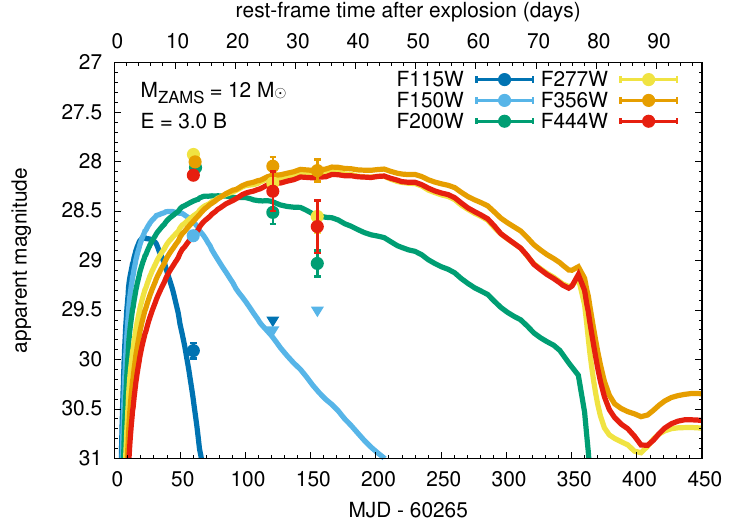} 
  \includegraphics[width=7.5cm]{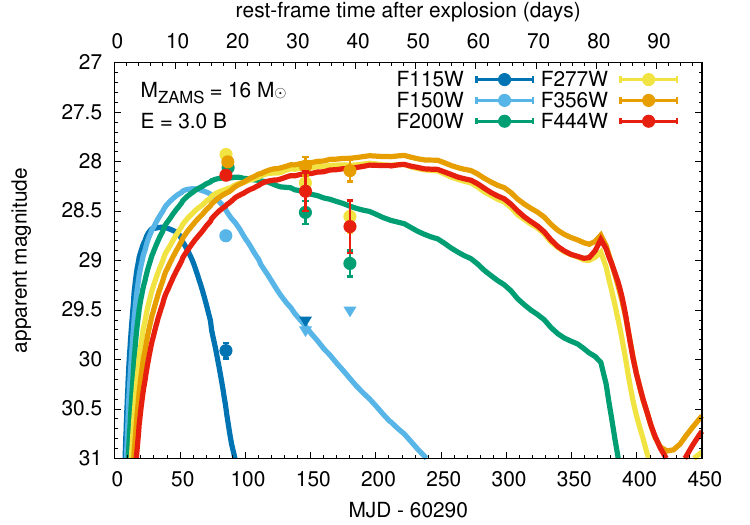} 
  \includegraphics[width=7.5cm]{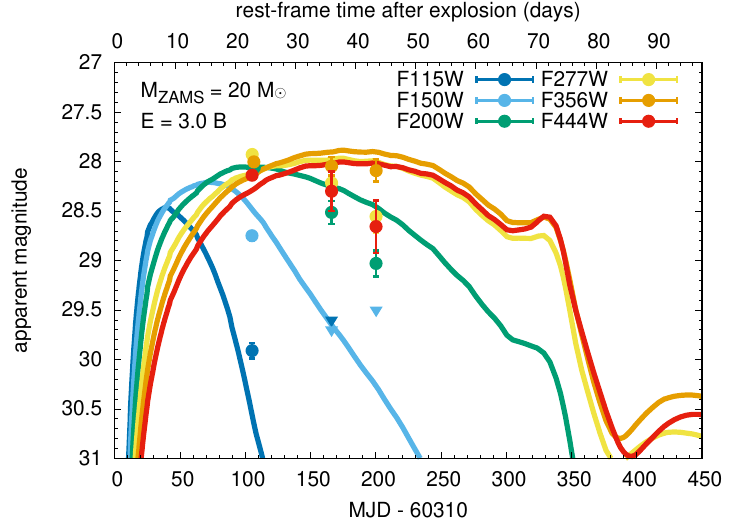} 
  \includegraphics[width=7.5cm]{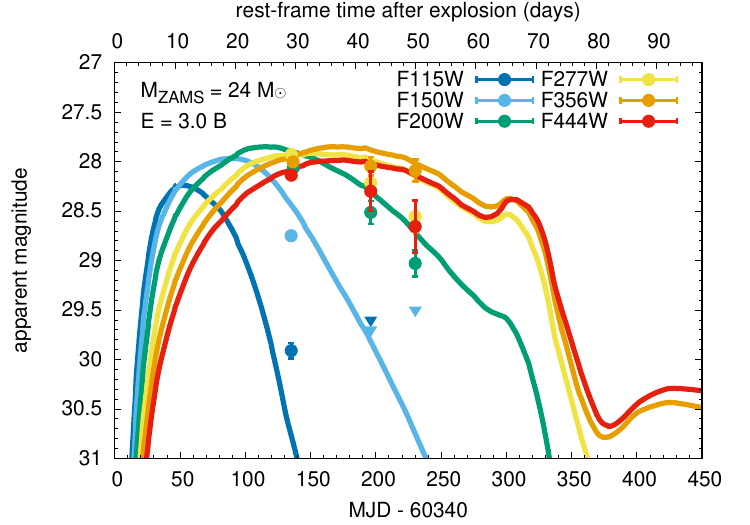} 
 \end{center}
\caption{Light curves of AT~2023adsv at $z=3.61$ compared to our synthetic light curves. No host galaxy extinction is assumed. Each panel shows synthetic light curves with different ZAMS masses. CSM is not included in the models presented in this figure.
{Alt text: Model lines and observational points. Lower x axis shows time in the observer frame from 0 to 450 days and upper x axis shows time in the rest-frame from 0 to 97.6 days. y axis shows apparent magnitude from 31 mag to 27 mag.} 
}\label{fig:adsv}
\end{figure}

\begin{figure}
 \begin{center}
  \includegraphics[width=7.5cm]{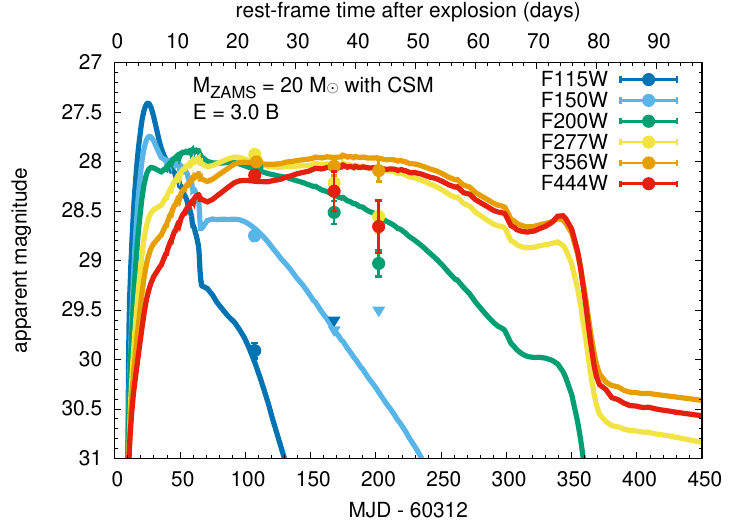} 
 \end{center}
\caption{Synthetic light curves with the confined dense CSM compared with AT~2023adsv. No host galaxy extinction is assumed. The CSM structure is from $\dot{M}=10^{-3}~\mathrm{M_\odot~yr^{-1}}$ with the terminal wind velocity of $10~\mathrm{km~s^{-1}}$. The radius of the confined CSM is $10^{15}~\mathrm{cm}$. The CSM mass is $0.23~\mathrm{M_\odot}$.
{Alt text: Model lines and observational points. Lower x axis shows time in the observer frame from 0 to 450 days and upper x axis shows time in the rest-frame from 0 to 97.6 days. y axis shows apparent magnitude from 31 mag to 27 mag.} 
}\label{fig:adsv_csm}
\end{figure}

\begin{figure}
 \begin{center}
  \includegraphics[width=7.5cm]{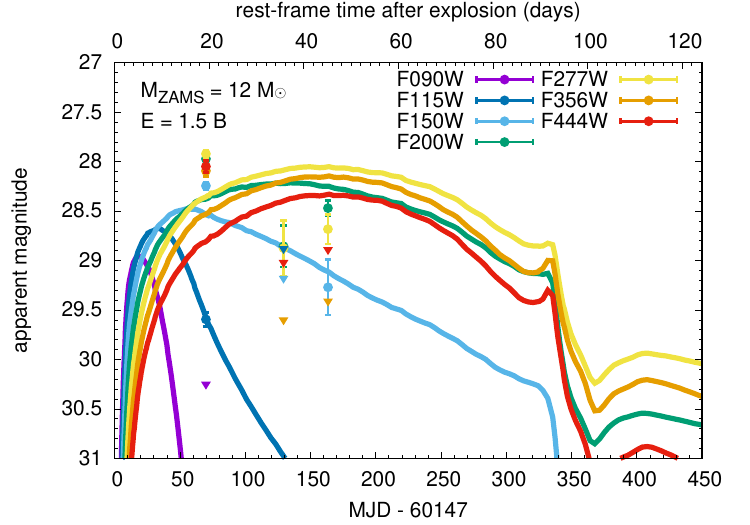} 
  \includegraphics[width=7.5cm]{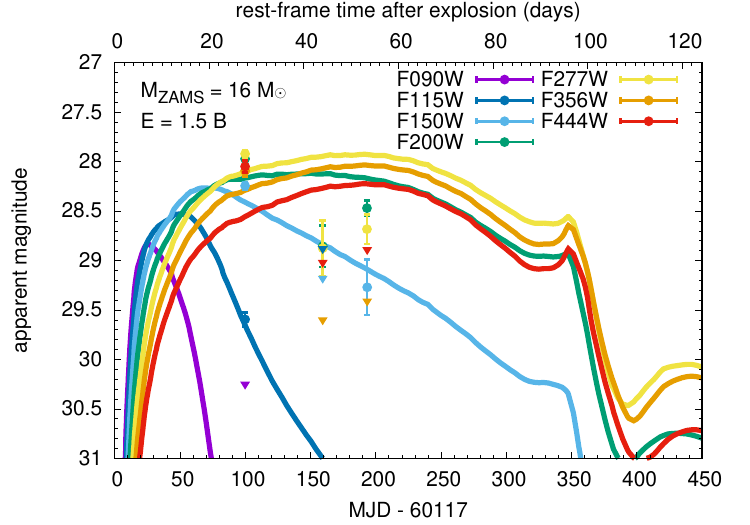} 
  \includegraphics[width=7.5cm]{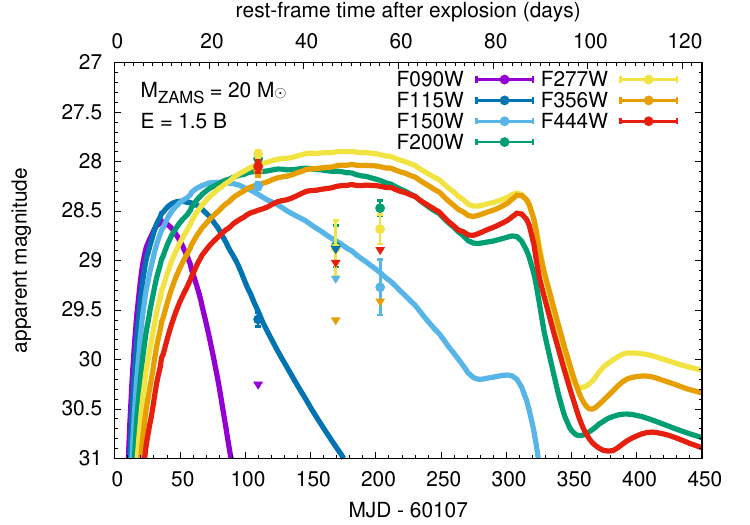} 
  \includegraphics[width=7.5cm]{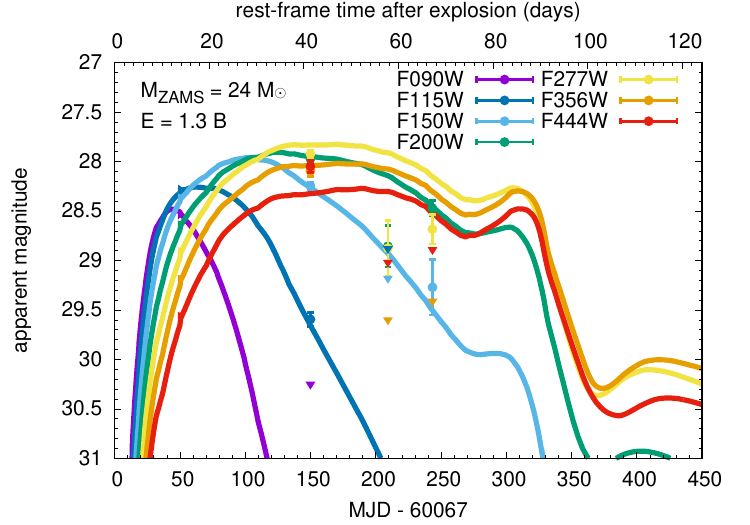} 
 \end{center}
\caption{Light curves of AT~2023adte at $z=2.623$ compared to synthetic light curves. No host galaxy extinction is assumed. Each panel shows synthetic light curves with different ZAMS masses. No CSM is attached to the progenitors in the models in this figure.
{Alt text: Model lines and observational points. Lower x axis shows time in the observer frame from 0 to 450 days and upper x axis shows time in the rest-frame from 0 to 124.2 days. y axis shows apparent magnitude from 31 mag to 27 mag.} 
}\label{fig:adte}
\end{figure}

\begin{figure}
 \begin{center}
  \includegraphics[width=7.5cm]{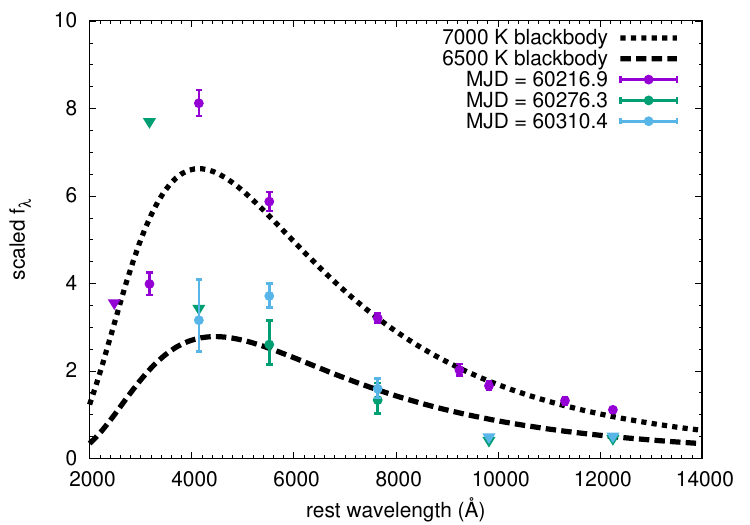} 
 \end{center}
\caption{SED evolution of AT~2023adte. The two dashed lines are the blackbody functions with $7000~\mathrm{K}$ and $6500~\mathrm{K}$.
{Alt text: Model lines and observational points. x axis is from 2000~\AA\ to 14000~\AA\ and y axis is from 0 to 10.} 
}\label{fig:adte_sed}
\end{figure}

\begin{figure}
 \begin{center}
  \includegraphics[width=7.5cm]{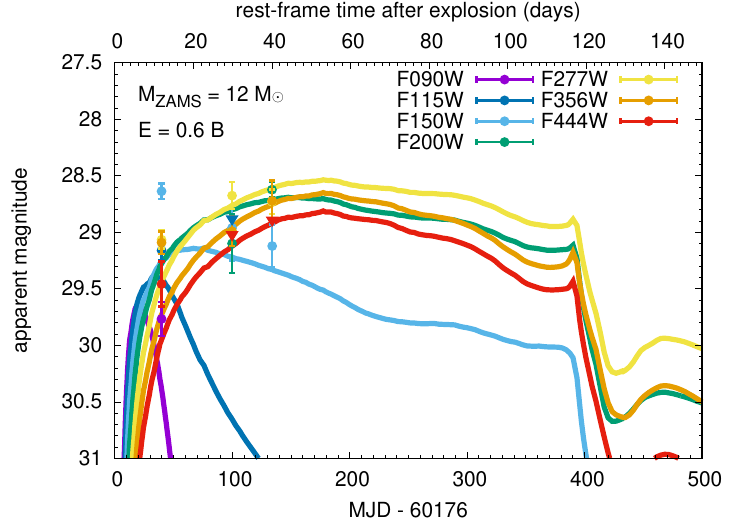} 
  \includegraphics[width=7.5cm]{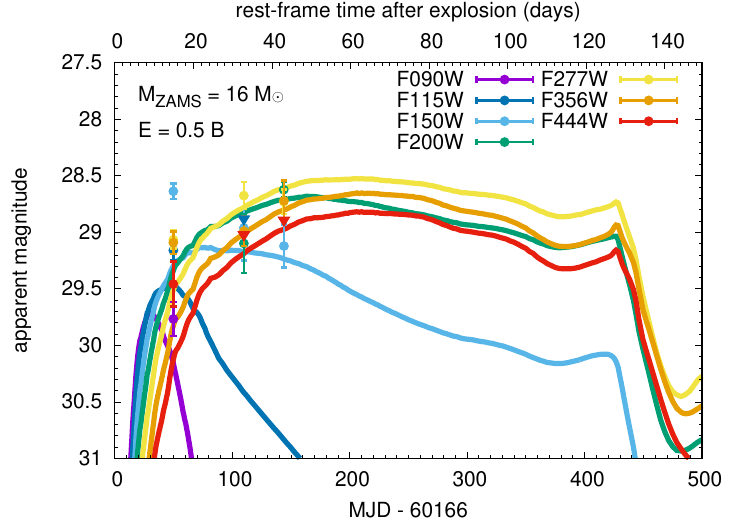} 
  \includegraphics[width=7.5cm]{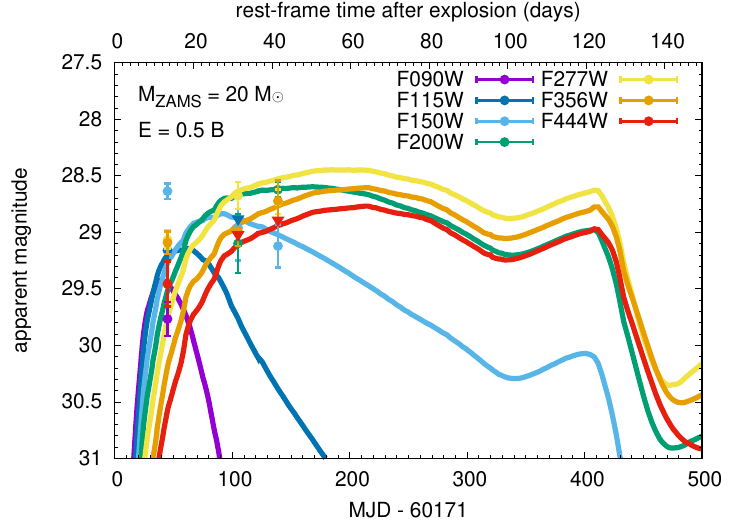} 
  \includegraphics[width=7.5cm]{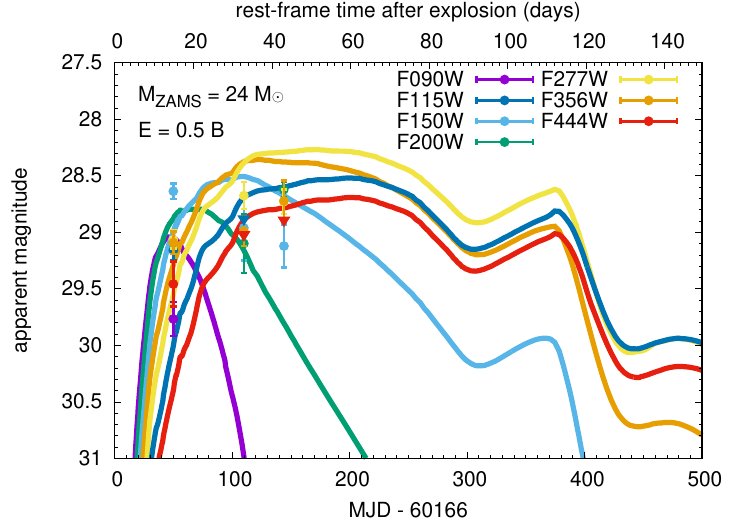} 
 \end{center}
\caption{Light curves of AT~2023adtf at $z=2.344$ compared to synthetic light curves without CSM. No host galaxy extinction is assumed. Each panel shows synthetic light curves with different ZAMS masses.
{Alt text: Model lines and observational points. Lower x axis shows time in the observer frame from 0 to 500 days and upper x axis shows time in the rest-frame from 0 to 149.5 days. y axis shows apparent magnitude from 31 mag to 27.5 mag.}
}\label{fig:adtf}
\end{figure}

\begin{figure}
 \begin{center}
  \includegraphics[width=7.5cm]{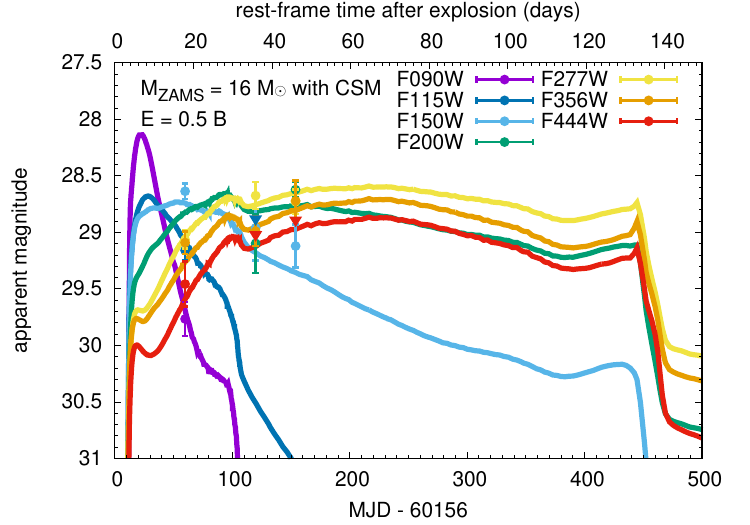} 
 \end{center}
\caption{Synthetic light curves of AT~2023adtf ($z=2.344$) with the confined dense CSM. No host galaxy extinction is assumed. The CSM structure is from $\dot{M}=1.3\times10^{-3}~\mathrm{M_\odot~yr^{-1}}$ with the terminal velocity of $10~\mathrm{km~s^{-1}}$. The radius of the confined CSM is $10^{15}~\mathrm{cm}$. The CSM mass is $0.23~\mathrm{M_\odot}$.
{Alt text: Model lines and observational points. Lower x axis shows time in the observer frame from 0 to 500 days and upper x axis shows time in the rest-frame from 0 to 149.5 days. y axis shows apparent magnitude from 31 mag to 27.5 mag.}
}\label{fig:adtf_csm}
\end{figure}

\begin{figure}
 \begin{center}
  \includegraphics[width=7.5cm]{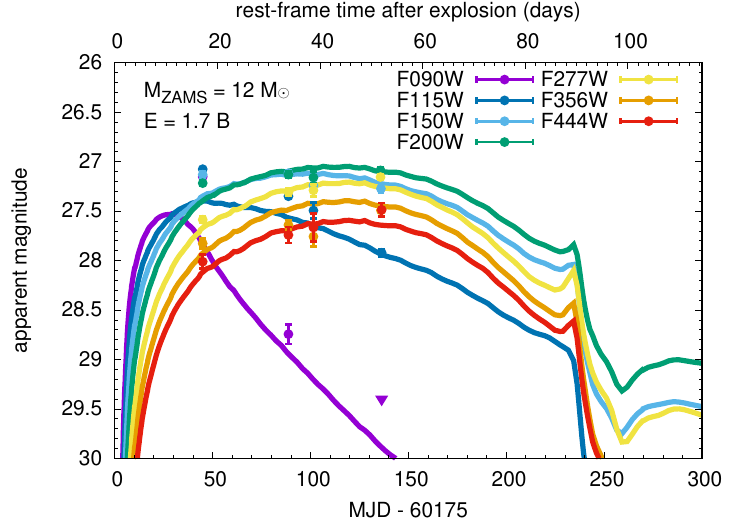} 
  \includegraphics[width=7.5cm]{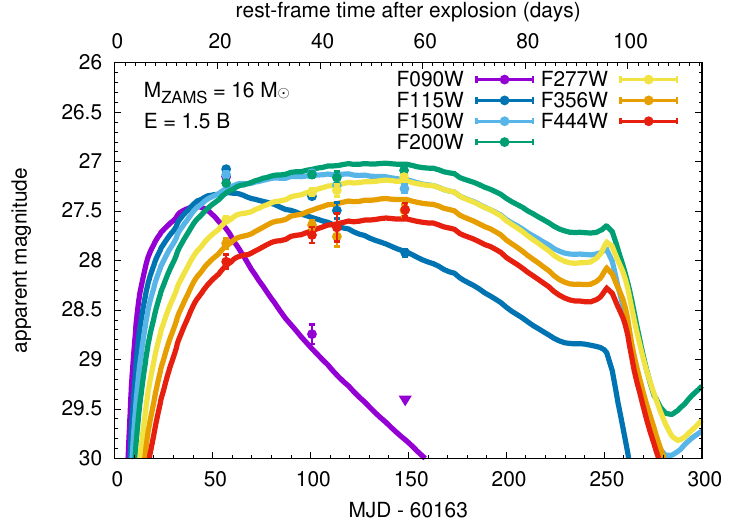} \\
  \includegraphics[width=7.5cm]{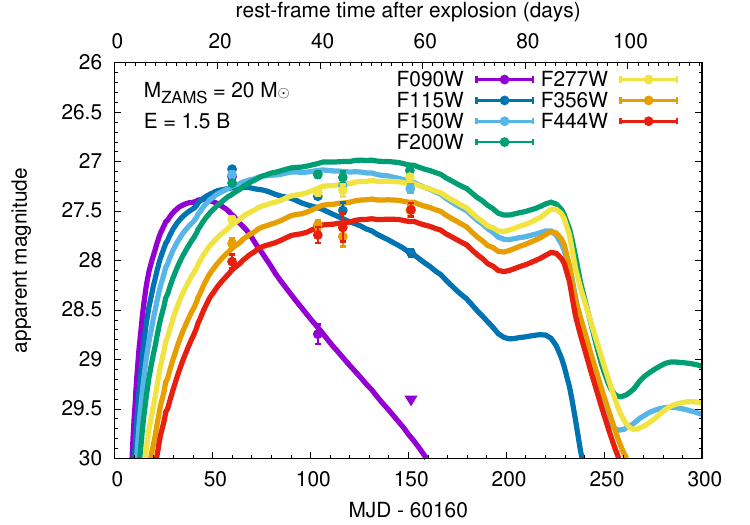} 
  \includegraphics[width=7.5cm]{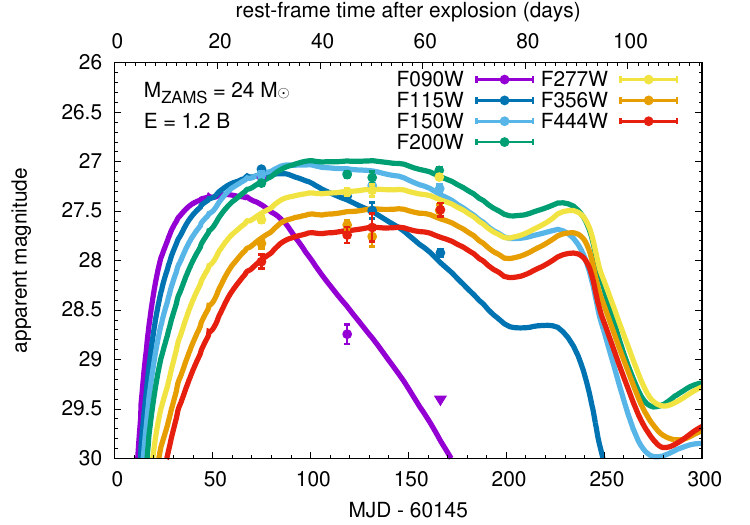} 
 \end{center}
\caption{Light curves of SN~2023adto at $z=1.62$ and synthetic light curves. No host galaxy extinction is assumed. No CSM is required to reproduce the light-curve properties during the observed period. Each panel shows synthetic light curves with different ZAMS masses.
{Alt text: Model lines and observational points. Lower x axis shows time in the observer frame from 0 to 300 days and upper x axis shows time in the rest-frame from 0 to 114.5 days. y axis shows apparent magnitude from 30 mag to 26 mag.}
}\label{fig:adto}
\end{figure}

\begin{figure}
 \begin{center}
  \includegraphics[width=7.5cm]{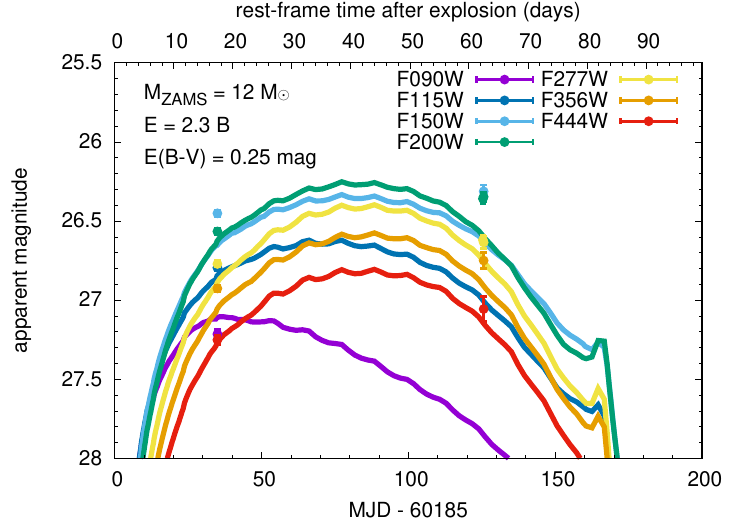} 
  \includegraphics[width=7.5cm]{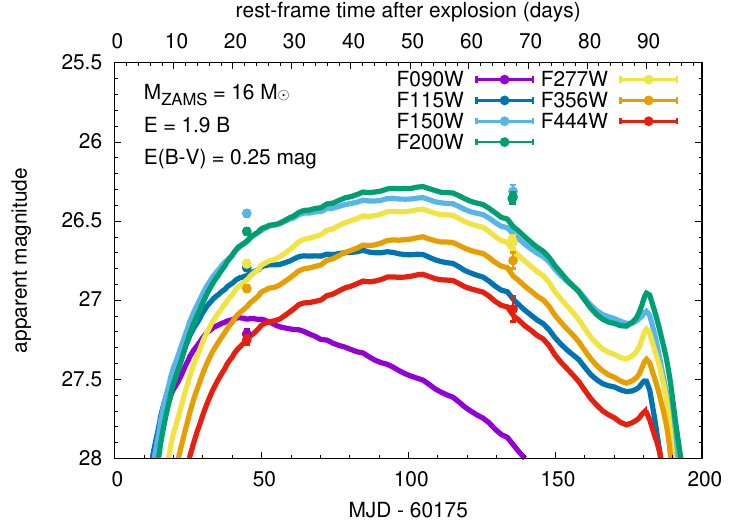} \\
  \includegraphics[width=7.5cm]{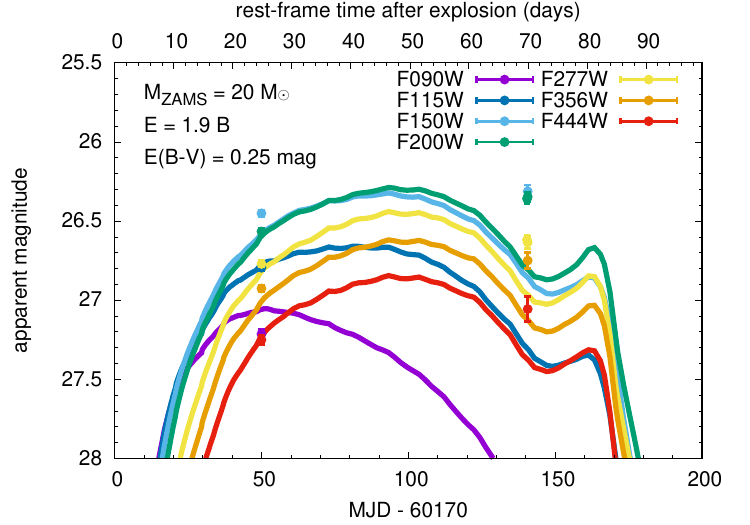} 
  \includegraphics[width=7.5cm]{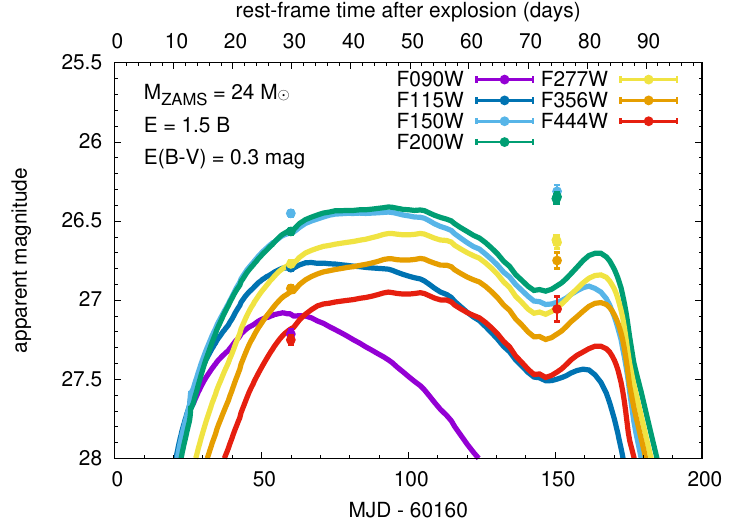} 
 \end{center}
\caption{Light curves of SN~2023adtu at $z=1.01$ and synthetic light curves. Each panel shows synthetic light curves with different ZAMS masses and the assumed host galaxy extinction is shown in each panel. No CSM is required to reproduce the light-curve properties during the observed period.
{Alt text: Model lines and observational points. Lower x axis shows time in the observer frame from 0 to 200 days and upper x axis shows time in the rest-frame from 0 to 99.5 days. y axis shows apparent magnitude from 28 mag to 25.5 mag.}
}\label{fig:adtu}
\end{figure}

\begin{figure}
 \begin{center}
  \includegraphics[width=7.5cm]{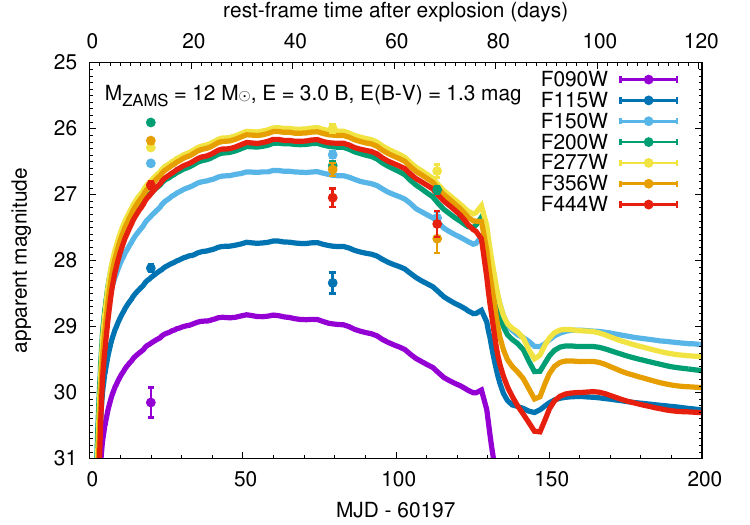} 
  \includegraphics[width=7.5cm]{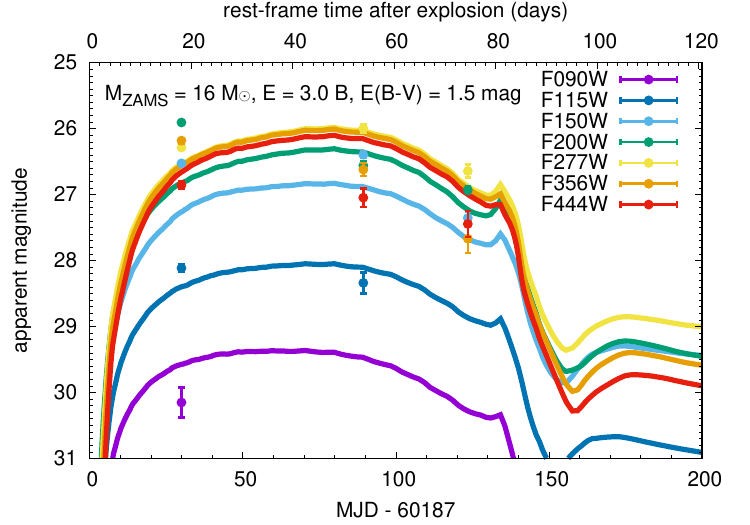} \\
  \includegraphics[width=7.5cm]{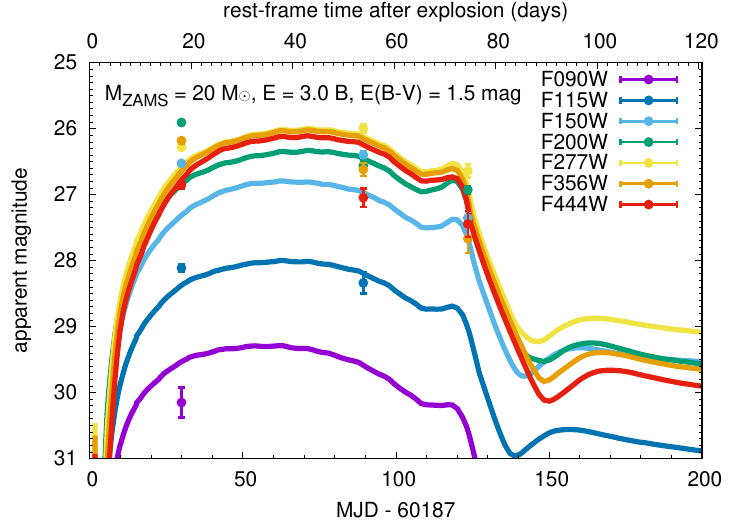} 
  \includegraphics[width=7.5cm]{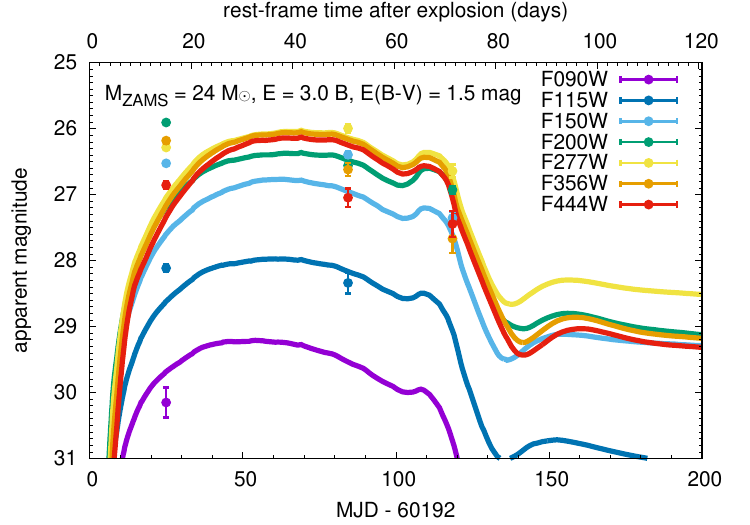} 
 \end{center}
\caption{
Light curves of AT~2023adtw at $z=0.657$ and synthetic light curves. Each panel shows synthetic light curves with different ZAMS masses and the assumed host galaxy extinction is shown in each panel. No CSM is attached in the models in this figure. The existence of the confined dense CSM does not improve the light-curve fits.
{Alt text: Model lines and observational points. Lower x axis shows time in the observer frame from 0 to 200 days and upper x axis shows time in the rest-frame from 0 to 120.7 days. y axis shows apparent magnitude from 31 mag to 25 mag.}
}\label{fig:adtw_lc}
\end{figure}

\begin{figure}
 \begin{center}
  \includegraphics[width=7.5cm]{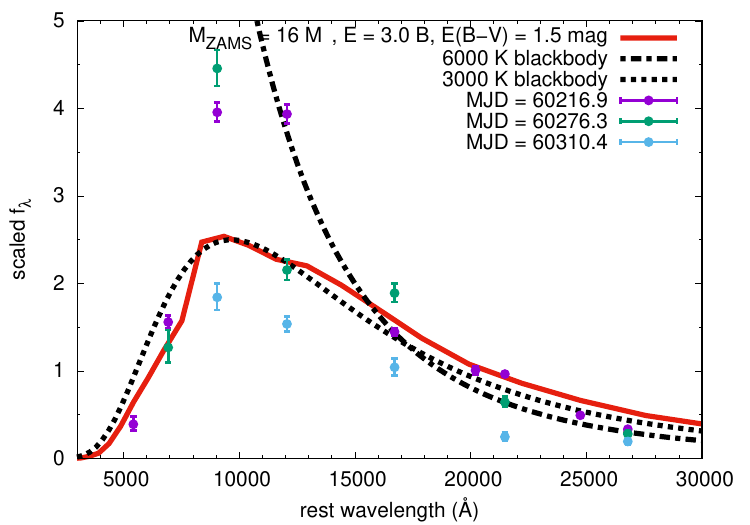} 
 \end{center}
\caption{
SED evolution of AT~2023adtw. The dot-dashed line shows the blackbody function for $6000~\mathrm{K}$ and the dashed line shows the blackbody function for $3000~\mathrm{K}$. The synthetic SED from the model of $\Mzams=16~\Msun$ and $E=3.0~\mathrm{B}$ with the host galaxy extinction of $E(B-V)=1.5~\mathrm{mag}$ at the first epoch ($\mathrm{MJD}=60216.9$) is presented for comparison.
{Alt text: Model lines and observational points. x axis is from 3000~\AA\ to 30000~\AA\ and y axis is from 0 to 5.} 
}\label{fig:adtw_sed}
\end{figure}

\subsection{Light-curve calculations}\label{sec:lightcurve}
The SN light-curve modeling is performed by using the one-dimensional multi-frequency radiation hydrodynamics code \texttt{STELLA} \citep{blinnikov1998,blinnikov2000,blinnikov2006}. \texttt{STELLA} was previously used for modeling light curves of zero metallicity Type~II SNe \citep{tolstov2016,moriya2019muller}. \texttt{STELLA} numerically follows the SED evolution at each time step. Thus, we can directly compare the observer-frame light curves obtained by JWST with the theoretical models by redshifting synthetic SEDs. The synthetic SEDs have 100 frequency bins ranging from $6.3\times 10^{13}~\mathrm{Hz}$ ($5\times 10^4$~\AA) to $2.8\times 10^{8}~\mathrm{Hz}$ ($1$~\AA) in a log scale. This wavelength binning is sufficient for light-curve modeling of Type~II SNe \citep{paxton2018}. We set the mass cut at $1.4~\Msun$ and artificially explode the progenitors above the mass cut. The explosion is initiated by inserting thermal energy above the mass cut without assuming any specific explosion mechanisms. We assume that 0.04~M$_{\odot}$ of $^{56}$Ni, which is a median \Ni mass estimated with local Type~II SN samples \citep[e.g.,][]{anderson2019,martinez2022}, exists uniformly within the helium core, but the \Ni decay heating does not affect our conclusions because all the Type~II SNe we investigate are only observed at luminous phases when the effects of the \Ni decay heating are insignificant. 

A significant fraction of Type~II SN light curves shortly after explosion are found to be affected by the interaction between SN ejecta and confined dense circumstellar matter (CSM, e.g., \citealt{forster2018,bruch2021}; \citealt{jacobson-galan2024}). As we discuss in the following sections, the light curves of some Type~II SNe in our sample can be better reproduced by taking the confined dense CSM into account. When we attach confined dense CSM, we adopt the following density structure \citep[e.g.,][]{moriya2017,moriya2018}.
\begin{equation}
    \rho_\mathrm{CSM}=\frac{\dot{M}}{4\pi v_\mathrm{wind}(r)}r^{-2},\label{eq:csm}
\end{equation}
where
\begin{equation}
    v_\mathrm{wind}(r) = v_0 +(v_\infty - v_0)\left(1-\frac{R}{r}\right)^\beta,
\end{equation}
$\dot{M}$ is a mass-loss rate, $v_0$ is a wind velocity at the progenitor surface, $v_\infty$ is a terminal wind velocity, and $R$ is a progenitor radius. In this work, we fix $\beta=2$, $v_\infty=10~\mathrm{km~s^{-1}}$, and a CSM radius of $10^{15}~\mathrm{cm}$. $\beta>2$ is observed in RSGs \citep[e.g.,][]{bennett2010} and $v_\infty=10~\mathrm{km~s^{-1}}$ is a typical wind velocity of RSGs \citep[e.g.,][]{goldman2017}. The confined dense CSM around Type~II SNe is often constrained to exist within $10^{15}~\mathrm{cm}$ \citep[e.g.,][]{yaron2017,silva-farfan2024}. $v_0$ is chosen so that the density structure of the progenitor and CSM is smoothly connected.

We also take the host galaxy extinction into account in modeling the light curves.
We apply the host galaxy extinction when the color of the SNe significantly deviates from that expected from the hydrogen recombination temperature (around $6000~\mathrm{K}$) so that the synthetic SEDs match the observed SEDs.
The host galaxy extinction is applied to the synthetic SEDs in the rest frame by assuming the extinction law of \citet{cardelli1989} with $R_V=3.1$ and then the extincted SEDs are redshifted. The Galactic extinction is ignored because no significant Galactic extinction exists towards the JADES Deep Field.

It is generally difficult to obtain models having perfect fits to all the observed photometry even for local well-observed Type~II SNe because of uncertainties in, e.g., opacity \citep[e.g.,][]{kozyreva2020}. Thus, we search for synthetic models that provide overall good matches to the observed photometry by eye as performed in estimating the properties of nearby Type~II SN \citep[e.g.,][]{kozyreva2022}. More sophisticated estimations using, e.g., Bayesian approaches, have been conducted for local Type~II SNe \citep[e.g.,][]{forster2018,subrayan2023,silva-farfan2024}. However, we do not take this approach because low-metallicity Type~II SN models required for this kind of methods are not available (see \citealt{moriya2023} for an example of solar-metallicity Type~II SN models of this kind). 
Because the last non-detection of our SN sample is one observer year before the explosion, their explosion epochs are not well determined. Thus, we treat them as a free parameter when comparing synthetic and observed light curves. Since light curves in bluer bands decline faster in Type~II SNe, the explosion time required for each progenitor model to reproduce the observed light curves are constrained reasonably well by combining all the available photometry and color evolution information across the rest-frame optical wavelength range.

\subsection{Host galaxy SED fitting}\label{sec:hostfitting}
We perform host galaxy SED fitting based on the JADES photometry obtained before 2023 which is not contaminated by SNe. We use these fits to estimate dust attenuation in the host galaxy to compare with the estimates derived from fitting the SN light-curves and discuss their differences, because the local attenuation of the SN site does not necessarily match the global attenuation derived from the host galaxy fit. For the fits, we use the SED fitting package \texttt{Prospector} \citep{leja2017, johnson2021}, and fit to the JADES NIRCam photometry in the F090W, F115W, F150W, F200W, F277W, F335M, F356W, F410M, and F444W filters. We supplement this photometry with NIRCam medium band data from both the First Reionization Epoch Spectroscopically Complete Observations \citep[FRESCO, ][]{Oesch2023} and from the JWST Extragalactic Medium Survey \citep[JEMS, ][]{Williams2023}: F182M, F210M, F430M, F460M, and F480M. Finally, at short wavelengths, we utilize photometry from Hubble Space Telescope (HST) Advanced Camera for Surveys (ACS) observations in these filters: F435W, F606W, F775W, F814W, and F850LP. 

For the \texttt{Prospector} fits, we follow Helton et al. (in preparation), and we refer to their thorough discussion of the fitting parameters. In brief, we used the Flexible Stellar Population Synthesis (FSPS) code \citep{conroy2009, conroy2010}, and we sampled the posterior distributions of the stellar population properties using the dynamic nested sampling code \texttt{dynesty} \citep{speagle2020}. We utilize a Chabrier initial mass function (IMF) with a lower bound of 0.08 $M_{\odot}$ and an upper bound of 120 $M_{\odot}$. Additionally, we assume a delayed-$\tau$ star-forming history of the form $\mathrm{SFR} \sim t_{\mathrm{age}} \times e^{-t_{\mathrm{age}} / \tau}$, where SFR is the star formation rate, $t_{\mathrm{age}}$ is the age of the galaxy, and $\tau$ is the e-folding time. For each SN host fit, we fix the redshift to the values presented in Table \ref{tab:sample}.  

For exploring the dust attenuation derived from photometry, we use the two-parameter model from \citet{charlotfall2000}, where the first parameter describes the diffuse dust optical depth and the second is the stellar birth-cloud dust optical depth. Additionally, we employ a flexible attenuation curve from \citet{noll2009}, and the strength of the ultraviolet bump is tied to the slope of the attenuation curve following \citet{kriekconroy2013}. We parameterize the dust attenuation for each source using $A_V$, the attenuation (in units of magnitudes) observed in the rest-frame \textit{V} band. We report the 16th, 50th, and 84th percentile $A_V$ values derived from the fits for each of the galaxies.

\section{Type~II SN properties}\label{sec:properties}
We present the results of our light-curve modeling and estimated properties of the six high-redshift Type~II SNe in this section.

\subsection{AT~2023adsv at $z=3.61$} 
AT~2023adsv is a likely Type~II SN studied by \citet{coulter2024} in detail. The spectrum of AT~2023adsv was obtained by the JWST Director's Discretionary Time (DDT) program \#6541 \citep{egami2023ddt}. The spectrum is dominated by the host galaxy emission lines and it is difficult to obtain the SN spectral type. However, the host spectrum allowed us to obtain a firm  redshift as well as its metallicity. AT~2023adsv is photometrically classified as a Type~II SN.

Figure~\ref{fig:adsv} compares the observed and synthetic light curves. The models with the explosion energy ($E$) of $3.0~\mathrm{B}$ ($1~\mathrm{B}\equiv 10^{51}~\mathrm{erg}$) reproduces the overall brightness evolution in multiple filters. The brightness at the first epoch matches better to the higher ZAMS mass models with $\Mzams\gtrsim 20~\Msun$. No host galaxy extinction is required to reproduced the observed light curves, which is consistent with low extinction estimated from the host galaxy SED fitting ($A_V = 0.15^{+0.11}_{-0.07}$). As discussed in \citet{coulter2024}, the rest-frame ultraviolet brightness in the early phase can be better explained by introducing confined dense CSM. We present the synthetic light curve with confined CSM in Figure~\ref{fig:adsv_csm}. Including CSM with $\dot{M}=10^{-3}~\mathrm{M_\odot~yr^{-1}}$ having a mass of $0.23~\Msun$ can make the overall light-curve fit better from the blue to red filters in the early phases.

In this work, we adopted the progenitors with $Z=0.1~\Zsun$ for all the SNe in the sample. However, the host galaxy metallicity of AT~2023adsv is estimated to be $0.3~\Zsun$ based on the host galaxy line emission \citep{coulter2024}. Thus, \citet{coulter2024} adopted the progenitor models with $Z=0.3~\Zsun$ and estimated the properties of AT~2023adsv based on them. \citet{coulter2024} also found that the host galaxy SED fitting shows a possibility that the host galaxy has a much lower metallicity. Despite of the differences in the metallicity, they found similar explosion properties to those estimated here. They found that $\Mzams \simeq 20~\Msun$ reproduce the light curves best. The estimated explosion energy in \citet{coulter2024} ($2-3~\mathrm{B}$) is consistent with the estimate in this work ($3.0~\mathrm{B}$). The CSM property is also consistent in the two works. The two studies indicate that the slight difference in the progenitor metallicity does not affect the estimated Type~II SN properties much.

\subsection{AT~2023adte at $z=2.623$} 
AT~2023adte is photometrically classified as a Type~II SN in \citet{decoursey2024}. The redshift of AT~2023adte ($z=2.623$) is based on the host galaxy spectrum obtained before the explosion by JADES \citep{deugenio2024}. Figure~\ref{fig:adte} presents the light curves of AT~2023adte and the comparisons with our light curve models. The models with the explosion energies of $1.3-1.5~\mathrm{B}$ can reproduce the overall properties of the observed light curve without assuming host galaxy extinction. The host galaxy SED fitting shows relatively high extinction ($A_V = 0.79^{+0.12}_{-0.14}$). However, the SN is not located in the brightest region in the host galaxy (0.13" from the host galaxy center corresponding to 1~kpc, Figure~\ref{fig:face}) and it is possible that the SN is less extincted. The observed epochs are during the plateau phase of AT~2023adte. Indeed, the SEDs in the observed epochs are consistent with blackbody temperatures of $7000 - 6500~\mathrm{K}$ matching the hydrogen recombination temperature and no extinction is required to explain the color evolution (Figure~\ref{fig:adte_sed}). The $12~\Msun$ model does not become as bright as observed in the major bands in the first epoch, while the overall luminosity evolution is matched by the models with $\Mzams = 16-24~\Msun$. However, we find that the synthetic light curves are brighter than observed in the redder bands in all the models. The light-curve evolution can be reproduced without assuming the existence of confined dense CSM because the color evolution is consistent with that of the Type~II SN plateau phase.

\subsection{AT~2023adtf at $z=2.344$} 
AT~2023adtf is photometrically classified as a Type~II SN \citep{decoursey2024}. The host galaxy redshift is observed to be $z=2.344$ based on the spectroscopic observations by JADES before explosion \citep{bunker2023}. Figure~\ref{fig:adtf} presents the light curves of AT~2023adtf. The second and third epoch observations are consistent with the synthetic models from all the adopted ZAMS masses ($12-24~\Msun$) with the explosion energies of $0.5-0.6~\mathrm{B}$. No host galaxy extinction is required to reproduce the light-curve evolution. The host galaxy SED fitting provides $A_V=0.44^{+0.17}_{-0.17}$, but the SN is located at 0.14" away from the central bright region corresponding to 1.2~kpc at $z=2.344$.

The synthetic light curves do not become as luminous as the observed light curves at the first epoch. This inconsistency is resolved by taking a confined dense CSM into account. Figure~\ref{fig:adtf_csm} presents an example of the light-curve model with the confined dense CSM. By including the CSM with $\dot{M}=1.3\times 10^{-3}~\Msunpyr$, we can reproduce the early light-curve properties of AT~2023adtf. The corresponding CSM mass is $0.23~\mathrm{M_\odot}$. Figure~\ref{fig:adtf_csm} shows an example with the $\Mzams=16~\Msun$, but the early light curves can be reproduced by the CSM interaction regardless of the ZAMS mass.

\subsection{SN~2023adto at $z=1.62$} 
SN~2023adto is a Type~II SN with spectroscopic confirmation (Egami et al. in preparation). The SN spectrum was obtained by the DDT program \#6541 \citep{egami2023ddt}. 
Figure~\ref{fig:adto} shows the comparison between the observed light curves and synthetic light curves. The overall light-curve properties can be reproduced by all the ZAMS masses and the ZAMS mass is not well constrained. The estimated explosion energies are in the range of $1.2-1.7~\mathrm{B}$. No host galaxy extinction is required to reproduce the properties of SN~2023adto.  The host galaxy SED fitting shows $A_V=0.38^{+0.05}_{-0.05}$, but the SN is far from the central region (0.24" from the host galaxy center, corresponding to 2~kpc). We do not require the confined dense CSM to reproduce the light curves in the observed period.

\subsection{SN~2023adtu at $z=1.01$} 
SN~2023adtu is another spectroscopically confirmed Type~II SN at $z=1.01$ (Egami et al. in preparation). The spectrum was obtained by the DDT program \#6541 \citep{egami2023ddt}.
The light curves are presented in Figure~\ref{fig:adtu}. The observed light curves are redder than the synthetic light curves having the recombination temperature of around 6000~K and introducing the host galaxy reddening of $E(B-V)=0.25-0.3~\mathrm{mag}$ provides better matches to the observed color. No apparent host galaxy exists at the location of SN~2023dtu. The nearest galaxy exists about 0.5" away towards north east from the SN location corresponding to the physical distance of 4~kpc, and its galaxy SED fitting provides $A_V=0.36^{+0.17}_{-0.18}$. They are consistent with having moderate extinctions.
The observed light-curve duration is better explained by the models with $\Mzams=12$ and $16~\Msun$. The estimated explosion energies are $1.9-2.3~\mathrm{B}$. The observed light curve are consistent with those at the plateau phase of Type~II SNe and we are not allowed to constrain the existence of a confined dense CSM.

\subsection{AT~2023adtw at $z=0.657$} 
AT~2023adtw is photometrically classified as a Type~II SN in \citet{decoursey2024}. The redshift ($z=0.657$) is obtained by the host galaxy spectrum which is available in \citet{momcheva2016}. The light curves are presented in Figure~\ref{fig:adtw_lc}. The peak of the SEDs is consistent with the blackbody temperature of $3000~\mathrm{K}$, which is much lower than the hydrogen recombination temperature of around $6000~\mathrm{K}$ (Figure~\ref{fig:adtw_sed}). Thus, AT~2023adtw is likely affected by strong extinction. Our synthetic light-curve models can reproduce the observed light curves when we assume the host galaxy extinction of $E(B-V)=1.3-1.5~\mathrm{mag}$. Indeed, the host galaxy SED fitting also provides a high extinction estimate ($A_V=1.87^{+0.08}_{-0.11}$). In addition, AT~2023adtw appeared from a region that is bright in infrared, and it is consistent with having an even larger dust extinction. The SEDs have significant flux excess in F090W and F115W in the first two epochs but its origin is not clear (Figure~\ref{fig:adtw_sed}). The explosion models with $3.0~\mathrm{B}$ can reproduce the plateau brightness with the required large extinction to fit the observed SEDs. The ZAMS mass is found to be difficult to constrain because the plateau duration can be reproduced by all the progenitor models we adopt.

The first epoch is found to be brighter than the synthetic light curves in several bands (Figs.~\ref{fig:adtw_lc} and \ref{fig:adtw_sed}). In AT~2023adsv and AT~2023adtf, adding confined dense CSM improved the match between the observations and models. However, in the case of AT~2023adtw, introducing confined dense CSM significantly increases the brightness especially in F090W and F115W, which does not improve the overall fit. Thus, we are not able to constrain the properties of confined dense CSM around AT~2023adtw.

\begin{figure}
 \begin{center}
  \includegraphics[width=7.5cm]{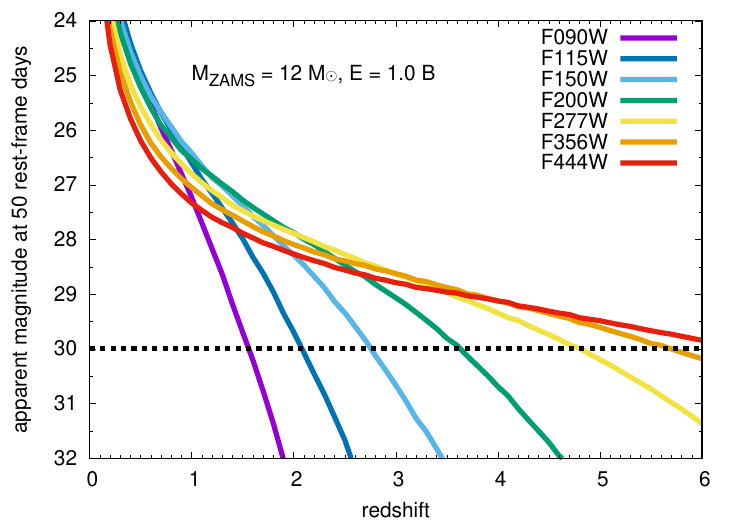} 
 \end{center}
\caption{
Apparent magnitudes of a standard Type~II SN model at around the middle of the plateau phase (50~days after the explosion in the rest frame). The standard model is obtained from the explosion of $\Mzams=12~\Msun$ with the explosion energy of $1~\mathrm{B}$. It has the absolute $V$ band magnitude of $-16.7~\mathrm{mag}$ at 50~days after the explosion in the rest frame with the plateau duration of around 100~days. The JADES transient survey limiting magnitude ($30~\mathrm{mag}$) is indicated by the dashed horizontal line.
{Alt text: Model lines. x axis from 0 to 6 and y axis from 32~mag to 24~mag.} 
}\label{fig:plateau_luminosity}
\end{figure}

\section{Discussion and conclusions}\label{sec:discussion}
We have estimated progenitor and explosion properties of six Type~II SNe at $0.675\leq z\leq 3.61$. The estimated properties are summarized in Table~\ref{tab:sample}. Because of the limited time coverage in the light curves, we do not obtain strong constraints on their ZAMS masses. If we assume explosion energies estimated from explosion simulations, the plateau brightness alone may constrain the ZAMS mass \citep{barker2022,barker2023}. However, it is not clear these explosion energies obtained by assuming the solar-metallicity progenitor models can be applied to the low-metallicity (around $0.1~\Zsun$) progenitor explosions and thus we treat explosion energies as a free parameter. Meanwhile, their explosion energies are constrained relatively well because the plateau brightness is strongly affected by the explosion energy \citep[e.g.,][]{popov1993,kasen2009}. The estimated explosion energies of Type~II SNe in the local Universe is mostly below $2~\mathrm{B}$ \citep[e.g.,][]{pumo2017,morozova2018,martinez2022,subrayan2023,silva-farfan2024}. However, we found two Type~II SNe (AT~2023adsv and AT~2023adtw) with the estimated explosion energy of $3.0~\mathrm{B}$ in our six Type~II SNe at high redshifts. The fraction of Type~II SNe with large explosion energies seems higher than those in the local Universe.

The seemingly high fraction of high energy events might originate from the observational biases to discover brighter events. Figure~\ref{fig:plateau_luminosity} presents the apparent magnitude of a typical Type~II SN model having $\Mzams=12~\Msun$ and $E=1~\mathrm{B}$ at the rest-frame 50~days from the explosion which is at the middle of the plateau phase. This model has the absolute \textit{V}-band magnitude of $-16.7~\mathrm{mag}$ with the plateau duration of around 100~days, which is typically observed in Type~II SNe \citep{anderson2014,anderson2024}. With the survey limiting magnitudes of around $30~\mathrm{mag}$, we should be able to discover standard Type~II SNe at $z>4$. One energetic Type~II SN, AT~2023adtw at $z=0.657$, is estimated to have a high extinction [$E(B-V)=1.3-1.5~\mathrm{mag}$ or $A_V = 4.0-4.7~\mathrm{mag}$]. SNe with such a high extinction were also discovered previously \citep[e.g.,][]{kool2018}. However, even if it were a standard event with a lower luminosity, it should have been above the detection limit (Figure~\ref{fig:plateau_luminosity}). Thus, it is possible that the JADES transient survey is not necessarily strongly biased to discover more energetic events at high redshifts. The explosion energies of Type~II SNe at high redshifts, which are expected to be in low metallicity environments, may therefore truly tend to be higher. Type~II SNe at lower metallicities may tend to have higher luminosities \citep{taddia2016,scott2019,tucker2024} and this fact may also be linked to possible higher explosion energies at lower metallicities. Some low-metallicity SN explosion models indeed tend to have high explosion energies \citep[e.g.,][]{muller2019}. However, \citet{gutierrez2018} and \citet{grayling2023} do not find a significant correlation between metallicity and Type~II SN luminosity. More high-redshift Type~II SNe are required to confirm their general explosion properties.

Confined dense CSM is found to exist commonly in nearby Type~II SNe \citep[e.g.,][]{forster2018,bruch2021}. Whether such a confined dense CSM exists across the cosmic history or not is key information to uncover the unknown mechanisms forming confined dense CSM. In addition, the interaction between the SN ejecta and confined dense CSM is suggested to be a production site of high energy cosmic rays and neutrinos \citep[e.g.,][]{murase2024,kimura2024}. Thus, whether confined dense CSM exists ubiquitously or not is also important information to understand the production sites of high energy cosmic rays and neutrinos across cosmic history. Among six high-redshift Type~II SNe in our sample, two Type~II SNe show the likely existence of confined dense CSM. Thus, at least some high-redshift Type~II SNe likely have confined dense CSM. The other four Type~II SNe were not observed early enough to constrain the existence of the confined dense CSM. We require long-term frequent observations of the same field to obtain well-sampled high-redshift Type~II SN light curves to constrain the properties of confined dense CSM at high redshifts.

Type~II SNe discovered by optical transient surveys in the local Universe do not usually have a high host-galaxy extinction \citep{dejaeger2018,forster2018,silva-farfan2024,subrayan2023}. It is also known that a certain fraction of nearby Type~II SNe are likely missed by optical transient surveys because of high dust extinction \citep[e.g.,][]{kool2018,fox2021}, and the fraction of highly extincted SNe can increase at higher redshifts \citep{mattilla2012}. In our six high-redshift Type~II SNe, four higher-redshift Type~II SNe at $1.62\leq z\leq 3.61$ discovered in the rest-frame optical wavelengths are consistent with having no host galaxy extinction. However, two lower-redshift Type~II SNe, SN~2023adtu at $z=1.01$ and SN~2023adtw at $z=0.657$, are estimated to have high extinction of $E(B-V)=0.25~\mathrm{mag}$ ($A_V = 0.8~\mathrm{mag}$) and $E(B-V)=1.3-1.5~\mathrm{mag}$ ($A_V = 4.0-4.7~\mathrm{mag}$), respectively. This result shows that we are able to discover both extincted and non-extincted SNe at various redshifts using JWST. More high-redshift Type~II SNe are required to explore their extinction distribution to see if Type~II SNe favor dusty environments at high redshifts.

We have discussed the high-redshift Type~II SN properties and their possible differences from the local Type~II SNe. However, because of our small sample size, it is still difficult to firmly conclude that there are possible differences among them. The long-term monitoring of the JADES field, as well as other fields such as COSMOS-Web \citep{casey2023,pierel2024cosmos}, is essential to discover more core-collapse SNe and to investigate any possible differences in the core-collapse SN properties in the current and early Universe.

\begin{ack}
Numerical computations were carried out on PC cluster at the Center for Computational Astrophysics, National Astronomical Observatory of Japan. This work is based on observations made with the NASA/ESA/CSA \textit{James Webb Space Telescope}. The data were obtained from the Mikulski Archive for Space Telescopes at the Space Telescope Science Institute, which is operated by the Association of Universities for Research in Astronomy, Inc., under NASA contract NAS 5-03127 for \textit{JWST}. These observations are associated with programs \#1180 and 6541.
The specific JWST observations analyzed can be accessed via \url{https://doi.org/10.17909/c4qk-xv53} (DOI: 10.17909/c4qk-xv53). 
JDRP is supported by NASA through an Einstein Fellowship grant No. HF2-51541.001 awarded by the Space Telescope Science Institute (STScI), which is operated by the Association of Universities for Research in Astronomy, Inc., for NASA, under contract NAS5-26555.
EE acknowledges the JWST/NIRCam contract to the University of Arizona NAS5-02015.
QW is supported by the Sagol Weizmann-MIT Bridge Program.
SA acknowledges support from the JWST Mid-Infrared Instrument (MIRI) Science Team Lead, grant 80NSSC18K0555, from NASA Goddard Space Flight Center to the University of Arizona.
BER acknowledges support from the NIRCam Science Team contract to the University of Arizona, NAS5-02015, and JWST Program 3215. The authors acknowledge use of the lux supercomputer at UC Santa Cruz, funded by NSF MRI grant AST 1828315.
\end{ack}

\section*{Funding}
TJM is supported by the Grants-in-Aid for Scientific Research of the Japan Society for the Promotion of Science (JP24K00682, JP24H01824, JP21H04997, JP24H00002, JP24H00027, JP24K00668) and by the Australian Research Council (ARC) through the ARC's Discovery Projects funding scheme (project DP240101786).
AJB acknowledges funding from the "FirstGalaxies" Advanced Grant from the European Research Council (ERC) under the European Union’s Horizon 2020 research and innovation programme (Grant agreement No. 789056).
ST acknowledges support by the Royal Society Research Grant G125142.

\section*{Data availability} 
The data underlying this article will be shared on reasonable request to the corresponding author.








\bibliographystyle{apj}
\bibliography{pasj}

\end{document}